 \documentclass[aps,pre,preprint,showpacs,showkeys,amsmath,amssymb,floatfix]{revtex4-1}

\usepackage{amsmath}    
\usepackage{amsfonts}   
\usepackage{amssymb}
\usepackage{graphicx}   
\usepackage{overpic}
\usepackage{bbm}

\usepackage{subcaption}

\newcommand{\ra}{\rightarrow}

\newcommand{\bfs}{{\bf s}}

\newcommand{\beq}{\begin{equation}}
\newcommand{\eeq}{\end{equation}}
\providecommand{\fx}{f_{\rm x_{\rm c}}}        
\begin{document}

\title{Finite-Size Effects on Return Interval Distributions for Weakest-Link-Scaling Systems}

\author{Dionissios T. Hristopulos}
 \email{dionisi@mred.tuc.gr}   
\author{Manolis P. Petrakis}
 \email{petrakis@mred.tuc.gr}   
 \affiliation{Department of Mineral Resources Engineering, Technical University of Crete,
Chania 73100, Greece}
\author{Giorgio Kaniadakis}
 \email{giorgio.kaniadakis@polito.it}
 \affiliation{Department of Applied Science and Technology, Politecnico di Torino,
Corso Duca degli Abruzzi 24, 10129 Torino, Italy}

\date{\today}

\begin{abstract}
The Weibull distribution is a commonly used model for the strength of brittle materials and
 earthquake return intervals. Deviations from Weibull scaling, however, have been observed in  earthquake return intervals
 and the fracture strength of quasi-brittle materials.
We investigate weakest-link scaling in finite-size systems
and deviations of empirical return interval distributions from the Weibull distribution function.
Our analysis employs the ansatz that the survival probability function of a system with complex interactions among its units  can be
expressed as the product of the survival probability functions for an ensemble of
representative  volume elements (RVEs). We show that
if the system  comprises a finite number of RVEs, it obeys the $\kappa$-Weibull distribution.
The upper tail of the $\kappa$-Weibull distribution declines as a power law in contrast with Weibull scaling.
The hazard rate function of the $\kappa$-Weibull  distribution
decreases linearly after a waiting time  $\tau_c \propto n^{1/m}$, where
$m$ is the Weibull modulus and $n$ is the system size in terms of representative  volume elements.
We conduct statistical analysis of
experimental data and simulations which shows  that the $\kappa$-Weibull provides competitive fits
to the return interval distributions of seismic data and
 of avalanches in a fiber bundle model.
In conclusion, using theoretical and statistical analysis of real and simulated data, we demonstrate
that the $\kappa$-Weibull distribution is a
useful model for  extreme-event return intervals in finite-size systems.
\end{abstract}



\maketitle

\section{Introduction}
Extreme events correspond to excursions of a random process $X(t)$, where $t$ is the time index,
to values  above or below a specified threshold  $z_q$. In natural processes, extreme events
include unusual weather patterns, ocean waves,  droughts, flash flooding, and earthquakes.
Such phenomena have important social, economic and ecological consequences.
The Fisher-Tippet-Gnedenko (FTG) theorem   states
 that if   $\{X_i\}_{i=1}^{n}$ are independent
and identically distributed (i.i.d.) variables, then a properly scaled affine
transformation of the minimum  $\chi_n := \min
(X_1, \ldots, X_n)$ follows asymptotically (for $n \ra \infty$) one of the
 extreme value  distributions, which include the Gumbel
(infinite support), reverse Weibull~\footnote{The term reverse Weibull
refers to the  distribution for maxima;
 this is known as the Weibull distribution in physical sciences and engineering.} (positive support)
and Fr\'{e}chet distributions (negative support)~\footnote{Counterpart distributions with reversed supports are obtained for maxima.}.
Whereas the FTG theorem is a valuable starting point, many  processes
of interest involve complex systems with correlated random variables. The impact of correlations on the
statistical behavior of complex physical systems thus needs to be  understood.
Early research on  extreme events statistics focused on purely statistical approaches~\cite{Gumbel35,Weibull51}.
Current efforts are based on nonlinear stochastic models and
aim to understand the patterns exhibited by extreme events and to control them~\cite{Sornette04a,Eliazar06,Franzke12,Nicolis12,Cavalcante13,Tall13}.

To improve risk assessment methodologies,  the statistics of the \emph{return intervals}, i.e., the time that elapses
between consecutive crossings of a given threshold by  $X(t)$, is an important property.
 If the threshold crossing implies failure  (e.g., fracture),
then the return intervals are intimately linked to the strength distribution of the
system~\cite{dth12}. Herein we  focus on the return intervals of earthquakes, i.e.,
\emph{earthquake return intervals} (ERI)~\footnote{The terms
\textit{interevent times, waiting times and recurrence
intervals} are also used.  A subtle distinction can
be made between \textit{recurrence intervals}, which refer to
events that take place on the same fault, and
\textit{interocurrence intervals} which encompass all faults in a
specified region~\cite{Abaimov08}. The statistical properties of recurrence
intervals are more difficult to estimate, because less information is available
for individual faults. The distinction is, nevertheless, conceptually important, since
recurrence intervals characterize the \emph{one-body}, i.e., the single-fault, problem, while
interoccurrence intervals are associated with the activity of the \emph{many-body system}~\cite{Sornette99}.
Herein we use the term return intervals without further distinction.}  and the return intervals of
avalanches in fiber bundle models under compressive loading.
From a broader  perspective, our scaling analysis can be also applied to other
systems or properties governed by weakest-link scaling laws, such as the strength
of quasibrittle heterogeneous materials.

This document is structured as follows. In the remainder of this section we review the literature on
 earthquake return intervals. Section~\ref{sec:weaklink} presents the basic principles of weakest-link scaling
 and its connection to the Weibull distribution. In Section~\ref{sec:wls-fin} we present an extension of weakest-link
 scaling for finite-size systems and motivate the use of the $\kappa$-Weibull distribution. Section~\ref{sec:wls-ri}
 links the $\kappa$-Weibull distribution  to earthquake return intervals using theoretical arguments.
 In Section~\ref{sec:eri} we apply these ideas to seismic data.
 Section~\ref{sec:fbm} focuses on the return intervals between avalanches in a fiber bundle model with global load sharing and demonstrates the performance of the  $\kappa$-Weibull distribution on this synthetic data.
 Finally, Section~\ref{sec:concl} summarizes our conclusions and briefly discusses the significance of the results.

\subsection{Earthquake Return Intervals}
Earthquake patterns can be investigated over different spatial supports which range
from a single fault to a system of faults~\cite{Sornette99}.
 Both isolated faults and fault systems represent complex problems that combine nonlinear and stochastic elements.
 Various probability functions have been proposed to model the earthquake return interval distribution (for a recent review see~\cite{Zhuang12}).
Several  authors have
proposed that earthquakes are manifestations of a self-organized system near a critical point~\cite{Bak02,Corral03,Saichev07}
or of a system near a spinodal critical point~\cite{Klein97,Serino11}.
Both cases imply the emergence of power laws.
Bak~{\it et al.}~\cite{Bak02} introduced a global scaling law that
relates earthquake return intervals with the magnitude and the distance
between the earthquake locations. These authors analyzed seismic catalogue data
over a period of 16 years  from an
extended area in California
 that includes several faults (ca. $3.35\times 10^5$ events).
 They observed power-law dependence over eight orders of
magnitude, indicating correlations over a wide range of return
intervals, distances and magnitudes. Corral and
coworkers~\cite{Corral03,Corral04,Corral06a,Corral06b} introduced a
local modification of the scaling law so that the return intervals
\textit{probability density function} (pdf) follows the universal expression $f_{\rm \tau}(\tau) \simeq \lambda
\tilde{f}(\lambda \, \tau)$, where $\tilde{f}( \tau)$ is a scaling
function and the typical return interval $\bar{\tau}$ is specific to
the region of interest.

Saichev and Sornette~\cite{Sornette06,Saichev07} generalized the scaling function
by  incorporating parameters with local dependence.
Their analysis was based on
the mean-field approximation of the return intervals pdf in the
epidemic-type aftershock sequence (ETAS) model~\cite{Ogata88}.
ETAS  incorporates  the main empirical laws of seismicity, such as
the Gutenberg-Richter dependence of earthquake frequency on
magnitude, the Omori-Utsu law for the rate of the aftershocks, and  a
similarity assumption that does not  distinguish
between foreshocks, main events and aftershocks (any event can
be considered as a trigger for subsequent events).

Several studies of earthquake catalogues and simulations show
that the \emph{Weibull distribution} is a good match for the empirical return intervals distribution~\cite{Hagiwara74,Rikitake76,Rikitake91,Sieh89,Yakovlev06,Turcotte06,Abaimov07,Abaimov08,Hasumi09a,Hasumi09b}.
In addition to statistical analysis, arguments supporting the {Weibull distribution} are based on
Extreme Value Theory~\cite{Santhanam08}, numerical simulations of slider-block models~\cite{Abaimov08}, and
 growth-decay models governed by the geometric Langevin equation~\cite{Eliazar06}.
The   \emph{Weibull distribution} is also used to model the fracture strength of
brittle and quasibrittle engineered materials~\cite{Hristopulos04,Bazant08,Bazant09}
and geologic media~\cite{Amaral08}.
With respect to extreme value theory, if we ignore correlations the FTG theorem favors the Weibull
  because the return intervals are
non-negative, whereas the Fr\'{e}chet distribution for minima has negative support and the Gumbel distribution has
unbounded support.

 A physical connection between the distribution of shear strength of the Earth's crust and the ERI distribution
  was proposed in~\cite{dth12}. According to a simplified stick-slip model, if the shear strength follows the
  Weibull distribution, under certain conditions the ERI also follows the Weibull distribution with parameters which are determined from the
 respective strength parameters and the exponent of the loading function.
The  conditions include: (i) the stress increase during the stick phase follows a power-law
function of time
(ii) the duration of the slip phase can be ignored (iii) the residual stress is uniform across
different stick-slip cycles, and (iv)
the parameters of the Earth's crust shear strength distribution are uniform over the study area.
 In particular, if the shear strength follows the
 Weibull distribution with \emph{modulus} $m_s$ and the stress increases  with time
 as a power law with exponent $\beta$ between consecutive events,
 then the ERIs  also follow the Weibull distribution with
 \emph{modulus} $m=m_{s}\beta$.
 On a similar track, a recent publication reports strong connections between the statistics of laboratory
 mechanical fracture experiments and earthquakes~\cite{Baro13,Main13}.

\section{Weakest-Link Scaling}
\label{sec:weaklink}

The weakest-link scaling   theory underlies the Weibull distribution.
Weakest-link scaling was founded by  the works of Gumbel~\cite{Gumbel35} and Weibull~\cite{Weibull51} on the statistics
of extreme values; it is used to  model the strength statistics of various disordered
materials~\cite{Curtin98,Hristopulos04,Alava06,Alava09}.
Weakest-link scaling  treats a disordered system as a chain of critical clusters, also known as
 links or representative volume elements (RVEs). The strength of the system
 is determined by the strength of the weakest link, hence the term weakest-link scaling ~\cite{Chakrabarti97}.
 The concept of links is straightforward in simple systems,
such as one-dimensional chains. In higher dimensions the RVEs correspond to
critical subsystems, possibly with their own internal structure,
failure of which destabilizes the entire system~\cite{Bazant06}.
We consider systems that follow weakest-link scaling and comprise $n$ links.
We use the symbol $x$ to denote the values of a random variable $X$ which can represent
mechanical strength or time intervals between two events.

We denote by $F_{1}^{(i)}(x)={\rm Prob}(X \le x)$ the \emph{cumulative distribution function (cdf)} that
$X$ takes values that do not exceed $x$. For example, if $X$ denotes mechanical strength (return intervals),
then $F_{1}^{(i)}(x)$ is the probability
 that the i-th link has failed when the loading has reached the value $x$ (when  time $\tau =x$ has passed).
Respectively, we denote by $F_{n}(x)$ the probability that the entire system fails at $x$.
The function  $R_{n}(x):=1- F_{n}(x)$ represents the system's \emph{survival probability}.
The principle of weakest-link scaling is equivalent to the statement that the
system's survival probability is equal to the product of the  link survival probabilities;
this is expressed mathematically as
\beq
\label{eq:Fs-system-1} R_{n}(x) = \prod_{i=1}^{n} R_{1}^{(i)}(x).
\eeq
If all the RVEs share the same functional form for $R_{1}^{(i)}(x)$,~\eqref{eq:Fs-system-1} leads to
\begin{equation}
\label{eq:Fs-system-2} R_{n}(x) =   \left[ R_{1}^{(i)}(x) \right]^n.
\end{equation}
Assuming that $R_{1}^{(i)}(x)$ is independent of $n$, Eq.~\eqref{eq:Fs-system-2} implies the following scaling
expression for $n>n'$
\begin{equation}
\label{eq:scale-F} R_{n}(x)= \left[ R_{n'}(x)
\right]^{n/n'}.
\end{equation}

If  the Weibull ansatz $R_{1}^{(i)}(x)={\rm e}^{-\phi(x)}$ is satisfied~\cite{Weibull51},
then $R_{n}(x) =  {\rm e}^{-n\,\phi(x)}.$
Furthermore, if $\phi(x) = (x/x_{0})^m$,  then $F(x) := F^{(i)}(x)$ for $i=1,\ldots, n$, and
\begin{equation}
\label{eq:Weib-cdf}
 F(x) = 1 - {\rm e}^{- \left( \frac{x}{x_{s}} \right)^{m}},
\end{equation}
where $x_s$ is the \emph{scale parameter} and $m>0$ is the
\emph{Weibull modulus or shape parameter}. The size dependence of $x_s$ is
determined by $x_s = x_0/n^{1/m}$~\footnote{We assume a fixed RVE size with volume $\propto n$.}.

Let us define the \emph{double logarithm of the inverse of the survival function} $\Phi_{n}(x) = \ln\ln R^{-1}_{n}(x)$.
In light of ~\eqref{eq:scale-F}, the following size-dependent scaling is obtained
\beq
\label{eq:Weib-scale}
\Phi_{n}(x) = \Phi_{n'}(x) + \ln(n/n').
\eeq
Based on the weakest-link scaling relation~\eqref{eq:Weib-scale} and the pioneering works~\cite{Daniels45,Smith81},
it can be shown  using asymptotic analysis that  the system's cdf
 tends asymptotically (as $n \ra \infty$) to the Weibull cdf~\cite{Smith81,Phoenix97}.
 Curtin then showed  that the large-scale cdf
parameters depend both on the system and the RVE size~\cite{Curtin98}.

The Weibull pdf is given by $f(x)=d F(x)/dx$ and leads to the expression
\beq
\label{eq:Weib-pdf}
f(x) = m\, \left( \frac{x}{\tau_{s}} \right)^{m-1} \, {\rm e}^{- \left( \frac{x}{x_{s}} \right)^{m}}.
\eeq
For $m<1$ the Weibull is also known as the \emph{stretched exponential distribution}~\cite{Sornette04a} and finds applications
in generalized relaxation models~\cite{ eliazar2012RARE,eliazar2013anomalous}, whereas for $m=2$ it is
equivalent to the \emph{Rayleigh distribution}. For $m<1$ the pdf has an integrable divergence at $x=0$ and  decays exponentially
 as $x \ra \infty$.  For $m=1$ the exponential pdf is obtained, whereas for $m>1$ the
pdf develops a single peak with diminishing width as $m \uparrow$.

Finally, for the Weibull distribution the function  $\Phi_{n}(x)$ is linearly related to the logarithm of $x$, i.e.,
$\Phi_{n}(x) =m \, \ln(x/x_s) $, and~\eqref{eq:Weib-scale} implies the size dependence
$\frac{{x}'_{s}}{{x}_{s}} = \left( \frac{n}{n'} \right)^{1/m}.$

\section{Weakest-link Scaling and Finite-size Systems}
\label{sec:wls-fin}
The  Weibull model assumes the existence of independent RVEs  and $n\gg 1$.
Nevertheless, there are systems for which the asymptotic assumption $n\gg 1$ is not \emph{a priori} justified.
For example, fault systems  span a  wide range of scales $(10^{0}$  - $10^{6}$ m).
The size or even the existence of an RVE are not established for fault systems.
In  quasibrittle materials, the RVE is assumed to exist but its size is not negligible compared to the
system size, leading to
deviations from the Weibull scaling in the upper tail
of the strength pdf~\cite{Bazant06,Bazant07,Bazant08,Bazant09}.
Using a piecewise Weibull-Gaussian model for the  RVE strength pdf, Bazant~{\it et al.}~\cite{Bazant06,Bazant07,Bazant08,Bazant09}
proposed that the system pdf exhibits a transition
from  Weibull scaling in the lower (left) tail to Gaussian dependence in the upper tail
at a probability threshold that moves upward as the size increases.

We consider a system that follows weakest-link scaling and
 consists of  RVEs with uniform properties. We associate the parameter
$\kappa$ with the number of effective RVEs through
$n=1/\kappa$. Hence, $\kappa$ (and also $n$) are parameters to be estimated from the data.
Note that $n$ does not need to be integer,
whereas for systems smaller than
one RVE  $n<1$ ($\kappa>1$) is possible.

\subsection{$\kappa$-Weibull Distribution}
\label{subsec:kweibull}

The exponential tail of the
Weibull pdf defined in~\eqref{eq:Weib-pdf} follows from the fact that the survival
probability $R(x)=\exp(-[x/x_s]^m)$ is defined in terms of the
exponential function $\exp(-z)$. On the other hand, in the last
decades particular attention has been devoted to pdfs that exhibit
power-law tails, namely $Ax^{-\alpha}$. Such dependence has been observed in
many branches of natural sciences including seismology, meteorology, and geophysics~\cite{eliazar2004growth,Kaniadakis09B}.

The simplest way to treat systems with these features is to replace the
exponential function in the definition of $R(x)$ by another proper
function which generalizes the exponential function and presents
power-law tails. A one-parameter generalization of the
exponential function has been proposed in~\cite{Kaniadakis01,Kaniadakis05} and is given by
\begin{equation}
    \label{eq:kappa}
    \exp_{\kappa}(z) = \left( \sqrt{1 +  {z^2}{\kappa^2} }  +
    {z}{\kappa}\right)^{1/\kappa},
\end{equation}
with $0 \leq \kappa <1$. The above generalization of the ordinary
exponential emerges naturally within the framework of special relativity, where the
parameter $\kappa$ is proportional to the reciprocal of light speed
~\cite{Kaniadakis09A,Lapenta08}. In that context, $\exp_{\kappa}(z)$ is the
relativistic generalization of the classical exponential $\exp(z)$.

The inverse function of the $\kappa$-exponential is the
\emph{$\kappa$-logarithm}, defined by
\begin{equation}
    \ln_{\kappa}(z) = \frac{z^{\kappa}-z^{-\kappa}}{2\kappa}.
    \label{eq:kappa_log}
\end{equation}

By direct inspection of the first few terms of the Taylor expansion
of $\exp_{\kappa}(z)$, reported in~\cite{Kaniadakis13}
\begin{equation}
    \label{eq:taylor}
    \exp_{\kappa}(z) = 1+z+\frac{z^2}{2}+ (1-\kappa^2)\frac{z^3}{3!} +
    (1-4\kappa^2)\frac{z^4}{4!} + \dots,
\end{equation}
it follows that when $z \rightarrow 0$ or $\kappa \rightarrow 0$  the
function  $\exp_{\kappa}(z)$ approaches the ordinary exponential i.e.
\begin{subequations}
\begin{equation}
\exp_{\kappa}(z) {\atop\stackrel{\textstyle\sim}{\scriptstyle 
z\rightarrow 0}} \exp(z), \label{eq:kappa_limxzero}
\end{equation}
\begin{equation}
\exp_{\kappa}(z) {\atop\stackrel{\textstyle\sim}{\scriptstyle
\kappa \rightarrow 0}} \exp(z).
  \label{eq:kappa_limkappazero}
\end{equation}
\end{subequations}
The most important feature of $\exp_{\kappa}(z)$ regards its
power-law asymptotic behavior~\cite{Kaniadakis05, Kaniadakis13} i.e.
\begin{equation}
\exp_{\kappa}(z) {\atop\stackrel{\textstyle\sim}{\scriptstyle
z\rightarrow \pm\infty}}\big|\,2\kappa z\big|^{\,\pm1/\kappa}.
  \label{eq:kappa_limxinf}
\end{equation}

We remark that the function $\exp_{\kappa}(-z)$ for $z \rightarrow 0$
coincides with the ordinary exponential i.e. $\exp_{\kappa}(-z)
\sim \exp(-z)$, whereas for $z \rightarrow +\infty$ it exhibits heavy tails
i.e.  $\exp_{\kappa}(-z) \sim (2\kappa z)^{\,-1/\kappa}$. Therefore the
function  $\exp_{\kappa}(-z)$ is particularly suitable to define the
survival probability~\cite{Clementi07,Clementi08}.  Following the change of variables
$\kappa=1/n$ and $z=(x/x_s)^m$ we obtain
\begin{equation}
R_{\kappa}(x)=\exp_{\kappa}\left ( -\left [ x/x_{s}\right ]^m \right ),
  \label{eq:kw_survival}
\end{equation}
The resulting $\kappa$-Weibull distribution exhibits a power-law tail
inherited by the $\kappa$-exponential:
\begin{equation}
F_{\kappa}(x)=1- \exp_{\kappa}\left ( -\left [ x/x_{s}\right ]^m \right), \label{eq:kw_cdf}
\end{equation}
\begin{equation}
f_{\kappa}(x)=\frac{m}{x_s}\left(\frac{x}{x_s}\right)^{m-1}
\frac{\exp_{\kappa}\left ( -\left [ x/x_{s}\right ]^{m} \right
)}{\sqrt{1+\kappa^2(x/x_s)^{2m}}}.
  \label{eq:kw_pdf}
\end{equation}

\begin{figure*}
\centering
\begin{subfigure}[b]{0.49\textwidth}
\includegraphics[width=\textwidth]{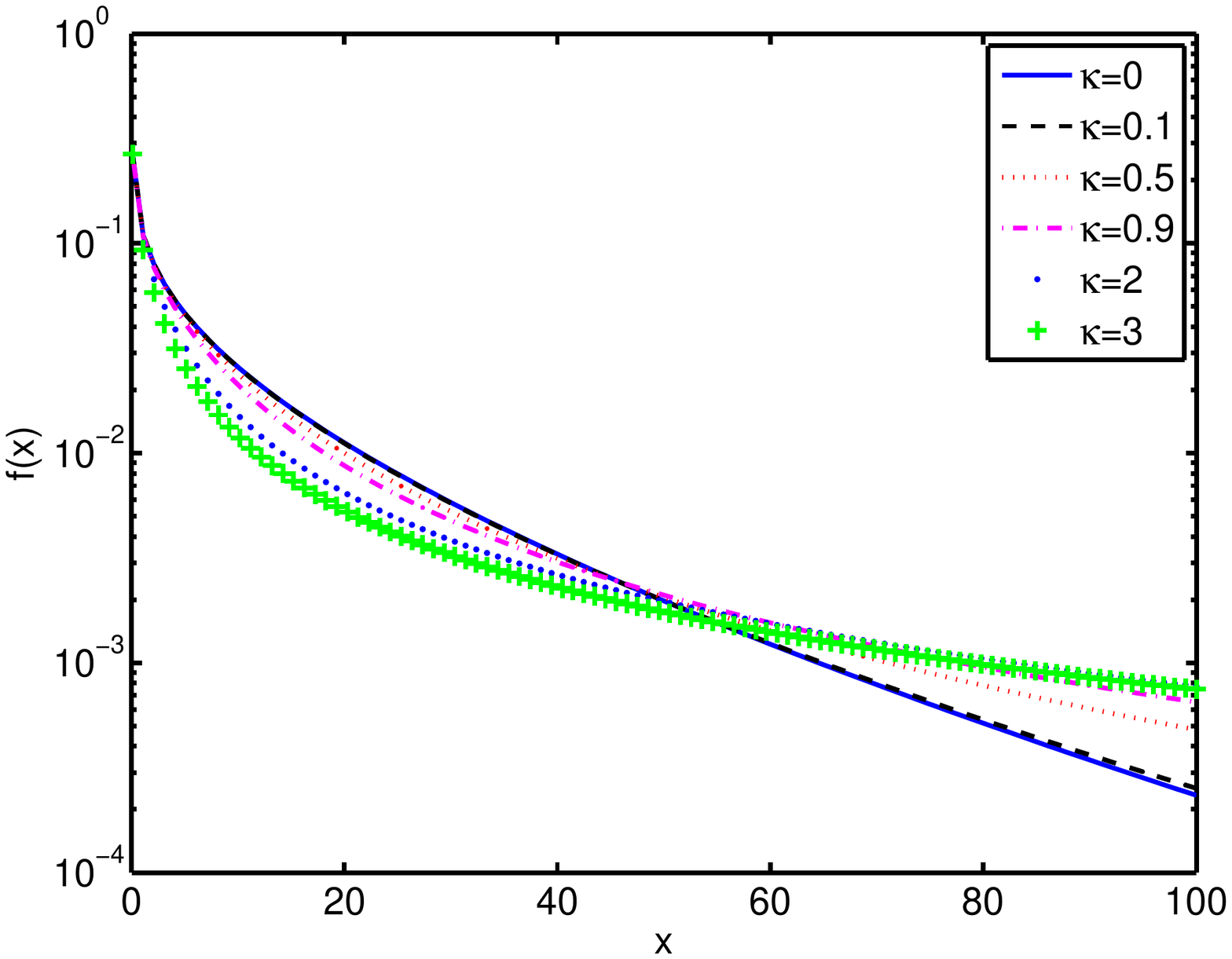}
\caption{$m=0.7$}
\label{fig:kpdf_1}
\end{subfigure}
\begin{subfigure}[b]{0.49\textwidth}
\includegraphics[width=\textwidth]{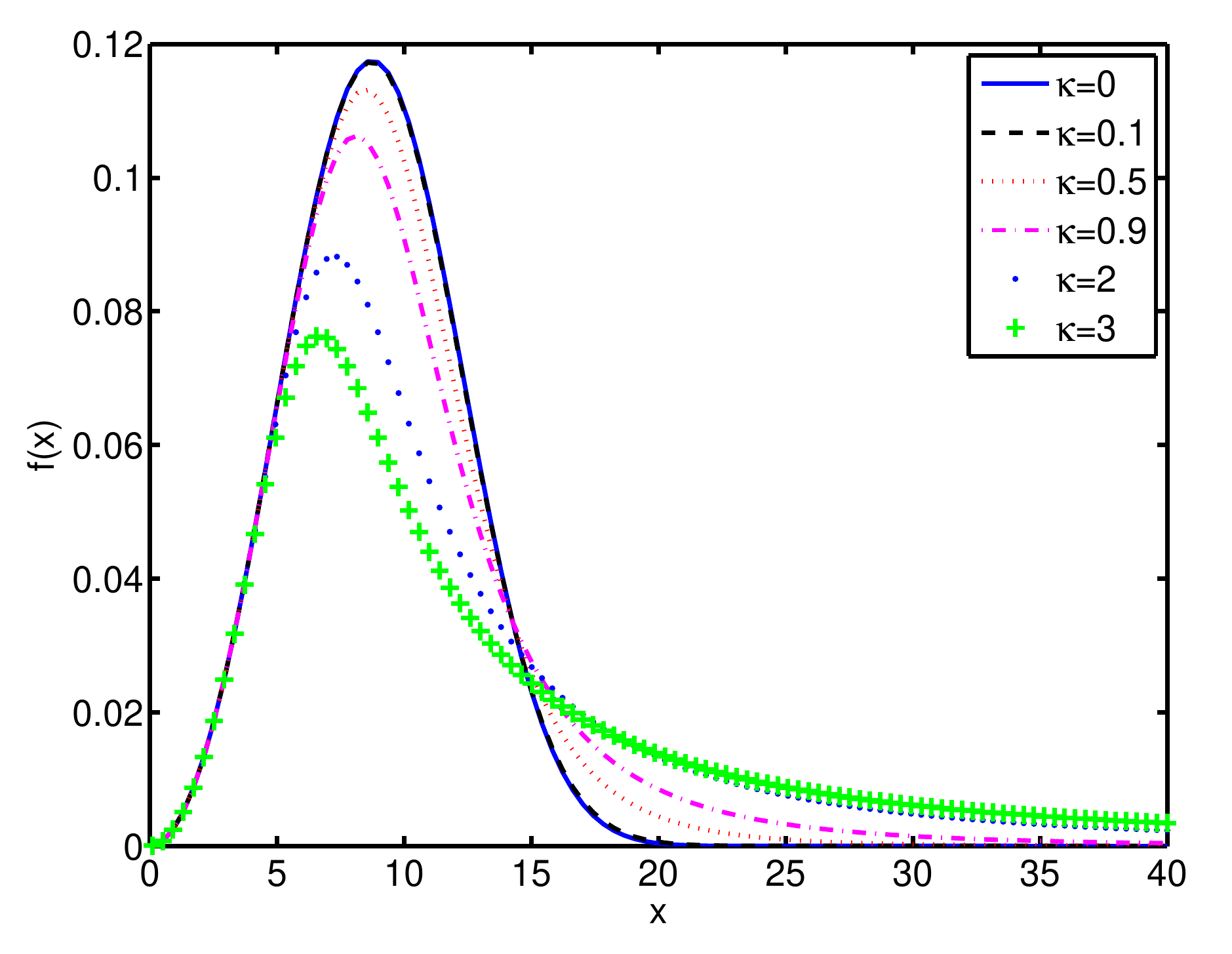}
\caption{$m=3$}
\label{fig:kpdf_2}
\end{subfigure}
\caption{\label{fig:kpdf} $\kappa$-Weibull pdfs for $x_{s}=10$, different values of $\kappa$, and (a) $m=0.7$ (b) $m=3$.}
\end{figure*}

Plots of the $\kappa$-Weibull  pdf for $x_{s}=10$, different $\kappa$, and two values of
$m$ ($m<1$ and $m>1$) are shown in Fig.~\ref{fig:kpdf}.
The plots also include the Weibull pdf $(\kappa=0)$ for comparison.
For both $m$ higher  $\kappa$ lead to a heavier right tail.
For $m=3$ the mode of the pdf moves to the left of $x_{s}$ as
$\kappa$ increases.
To the right of the mode, lower $\kappa$ correspond, at first, to
higher pdf values. This is reversed at a crossover point beyond which
the higher-$\kappa$ pdfs exhibit slower power-law decay for $x
\rightarrow \infty$, i.e., $f_{\kappa}(x) \propto x^{-\alpha}$, where
$\alpha = 1+ m/\kappa$.
The crossover point occurs at $\approx 1.5 x_{s}$ for $m=3$,
whereas for $m=0.7$ at $ \approx 5 \,x_{s}$.
For $m=0.7$ the mode is at zero independently of $\kappa$,
since the distribution is zero-modal for $m \le 1$.

It is important to note that the $\kappa$-Weibull admits explicit
expressions for
all the important univariate probability functions.
The $\kappa$-Weibull \emph{hazard rate function} is defined by means of
$h_{\kappa}(x)=f_{\kappa}(x)/R_{\kappa}(x)=- d\, \ln R_{\kappa}(x)/d x$, leading to
\begin{equation}
h_{\kappa}(x)=\frac{m}{x_s} \frac{\left(
x/x_{s}\right)^{m-1}}{\sqrt{1+\kappa^2(x/x_s)^{2m}}}. \label{eq:kw_hazard}
\end{equation}
The $\kappa$-Weibull \emph{quantile function} for a given survival probability $r$
is defined by
\begin{equation}
T_{\kappa}(r)=\frac{1}{x_s} \left(
\ln_{\kappa}{\frac{1}{r}}\right)^{1/m}.
\label{eq:kw_quantile}
\end{equation}
In addition,  if we
define  ${\Phi'}_{\kappa}(x)=\ln \ln_{\kappa} (1/R_{\kappa}(x))$ it
follows that
${\Phi'}_{\kappa}(x)= m \ln \left(\frac{x}{x_s}\right)$.
Hence, ${\Phi'}_{\kappa}(x)$
is independent of $\kappa$ and regains the logarithmic scaling of the
double logarithm of the inverse survival function.

\subsection{RVE Survival Function}
We   define the RVE cdf at level $x \in [0, \infty)$ through the equation
\beq
\label{eq:F1kappa}
F_{1}(x) = 1 + \frac{1}{n} \left( \frac{x}{x_{s}} \right)^{m} - \sqrt{1 +
    \frac{1}{n^2}\left(\frac{x}{x_{s}}\right)^{2m}}.
\eeq
 $F_{1}(x)$ is a well defined cdf, because  $F_{1}(x=0)= 0$, whereas for $x>0$
$F_{1}(x)$ is an increasing function of $x$, and $\lim_{x \ra \infty} F_{1}(x) = 1.$
This particular form of  $F_{1}(x)$ is motivated by arguments similar to those used in the
Weibull case.  In Section~\ref{sec:weaklink}, the Weibull survival function  $R_{n}(x)$
was derived from~\eqref{eq:Fs-system-1} assuming that the link survival function is
$R^{(i)}_{1}(x) =  {\rm e}^{-\phi(x)}$ where $\phi(x) = (x/x_{s})^m.$
Another approach that does not require the exponential dependence of the RVE
survival function is based on the following approximation
\[
R_{n}(x) = \left[ 1 - F_{1}(x)   \right]^{n} \Rightarrow \ln R_{n}(x) = n\, \ln\left[ 1 - F_{1}(x)   \right] \approx - n\,F_{1}(x).
\]
The above assumes that $F_{1}(x) \ll 1$ for the link cdfs if $n$ is large.
Then, assuming that $F_{1}(x) \propto (x/x_{s})^m$ the Weibull form is obtained.
The dependence of $F_{1}(x)$ for large $x$ which becomes relevant for finite $n$, however,
is not specified. In contrast,~\eqref{eq:F1kappa} generalizes the algebraic dependence so that $F_{1}(x) \sim (x/x_{s})^m$ for $x \ra 0$, whereas  $F_{1}(x)$ is
also well defined for $x \ra \infty$.

From~\eqref{eq:F1kappa} it follows that the respective survival function is
\beq
\label{eq:R1kappa}
R_{1}(x)= \sqrt{1 +
    \frac{1}{n^2}\left(\frac{x}{x_{s}}\right)^{2m}} - \frac{1}{n} \left( \frac{x}{x_{s}} \right)^{m}.
\eeq

Application of the weakest-link scaling relation~\eqref{eq:Fs-system-2} to~\eqref{eq:R1kappa}
leads to the following system survival function
\beq
\label{eq:Rnkappa}
R_{n}(x)= \left[\sqrt{1 +  \frac{1}{n^2}\left(\frac{x}{x_{s}}\right)^{2m}}
- \frac{1}{n} \left( \frac{x}{x_{s}} \right)^{m}\right]^{n}.
\eeq

The definition~\eqref{eq:F1kappa}  implies that
$F_{1}(x)$ and $R_{1}(x)$  depend on the number of RVEs,
which destroys the weakest-link scaling relation~\eqref{eq:scale-F}.
Based on~\eqref{eq:R1kappa} and using $z=(x/x_{s})^m$ it follows that
\[\frac{\partial R_{1}(z)}{\partial n}= \frac{z}{n^2} \left(1 - \frac{z}{\sqrt{z^2+n^2}} \right) >0,\; \forall z > 0.
\]
Hence, the survival probability of single RVEs at a given threshold $z$ increases with $n$. 

We propose an ansatz which is consistent with the dependence of $R_{1}(x)$ as given by~\eqref{eq:R1kappa}.
Assume that the system
comprises a number of units (e.g., faults) with inter-dependent RVE survival probabilities,  as expected
in the presence of correlations. Following a renormalization group (RG) procedure, the interacting units are replaced by non-interacting ``effective RVEs''. The RG procedure recovers the product form~\eqref{eq:Fs-system-2} for the
survival probability of independent RVEs, while renormalizing  the  scale parameter $x_0$ by
 the number of effective RVEs. We can think of $\kappa=1/n$ as a measure of the range of
interactions versus the size of the system; $\kappa=0$ yields the classical Weibull pdf for infinite systems,
whereas $\kappa \uparrow$
implies that the range of correlations increases thus reducing the number of
independent units; the case $\kappa=1$ means that the system can not be reduced to smaller independent units.

\section{Weakest-Link Scaling and Return Intervals}
\label{sec:wls-ri}

Below we focus explicitly on earthquake return intervals; thus, we replace $x$ with $\tau$.
In earthquake analysis the spatial support  includes either a single fault or a system of
several faults. The notion of an RVE with respect to earthquakes is neither theoretically developed
nor experimentally validated. Hence, herein we assume that the study domain involves $n $
\emph{independent, identically distributed}
RVEs, where $n$ is not necessarily an integer~\footnote{The number of RVEs $n$ may also depend on the earthquake cutoff magnitude.}.

\begin{figure}
    \centering
    \includegraphics[width=0.5\linewidth]{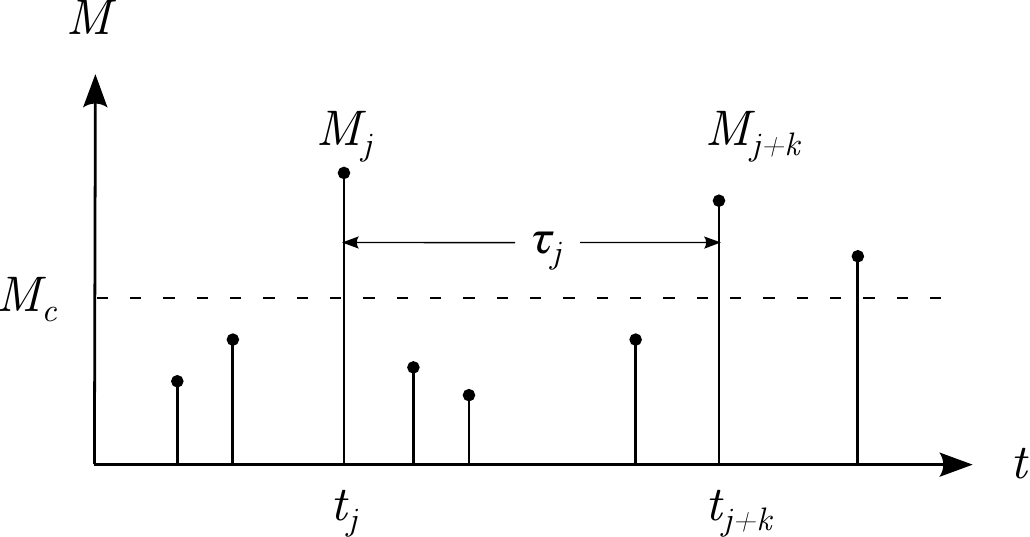}
    \caption{Schematic illustrating the definition of a return interval $\tau_j$ as the time between two consecutive events with magnitudes \ $M_j$ and $M_{j+k}$
     that exceed the threshold $M_c$, i.e., the events occurring at
    times $t_j$ and $t_{j+k}$.}
    \label{fig:ERI}
\end{figure}

An earthquake catalog is a table
 of the \emph{marked point process}~\cite{Daley03}
$C=\{{\bfs}_i, t_i, M_i\}_{i=1,\ldots,N}$, where ${\bfs}_i$ is the location, $t_i$ the time,
and $M_i$ the magnitude of the seismic event.
Given a threshold magnitude  $M_c$, an ERI sequence  comprises the intervals
$\{ \tau_{j} = t_{j+k} - t_{j} : (M_{j}, M_{j+k} > M_c) \wedge (M_{j+1}, \ldots, M_{j+k-1} \le M_c) \}$, where $j=1,\ldots, N_{c}-1,$ $N_{c}$ is the
number of events with magnitude exceeding $M_c$ (Fig.~\ref{fig:ERI}), and
$\wedge$ is the logical conjunction symbol.
The random variable $T_{M_{c}}^{(i)}$ $(i=1,\ldots,n)$ denotes the quiescent interval for the i-th RVE
during which no events of magnitude  $ M > M_c$  occur.
The cdf $F_{1}(\tau;M_c)={\rm Prob}(T_{M_{c}}^{(i)}\le\tau)$
 represents the probability of RVE
``failure'', i.e., that an event with $M > M_c$ occurs on the RVE
within time interval $\tau$ from the previous event.  In the following,
we suppress the dependence on $M_c$ for brevity.

\subsection{Survival Probability Function}

The \emph{survival  probability}
$R_{1}(\tau)= 1- F_{1}(\tau)$ is  the probability that no event with magnitude $M > M_c$
occurs on the RVE during the interval $T_{M_{c}}^{(i)} \le \tau$.
For  $\tau \ra 0$, it follows from~\eqref{eq:R1kappa} that $R_{1}(\tau) \sim 1- \tau^m/( n\,\tau_{s}^{m}) $.
For  $\tau \gg 1$ and finite $n$, it follows that $R_{1}(\tau) {\sim} n\,\tau_{s}^{m}/{2\tau^m}$,
and thus  $R_{n}(\tau)$  shows  the power-law dependence
$  R_{n}(\tau) {\sim} (n/2)^{n}(\tau_s /\tau)^{m n} $, characteristic of the \emph{Pareto distribution}.
In addition, $\lim_{n \ra \infty} R_{n}(\tau) = \exp \left[-   (\tau/\tau_s)^m\right]$,
thus recovering the Weibull survival probability at the limit of an infinite system.
The above equation shows that
 the \emph{interval scale} for large $n$ saturates at $\tau_s$, in contrast with
the classical $\tau_s \propto n^{-1/m}$ Weibull scaling.
Based on the Gutenberg-Richter law of seismicity which predicts exponential decay of earthquake events as $M_c \uparrow$, it follows that
   $\tau_s \uparrow$  as $M_c \uparrow$. In contrast, $m$ is expected to vary more slowly with $M_c$~\cite{dth12}.

\subsection{Median of Return Intervals}
\begin{figure}
\includegraphics[width=0.95\linewidth]{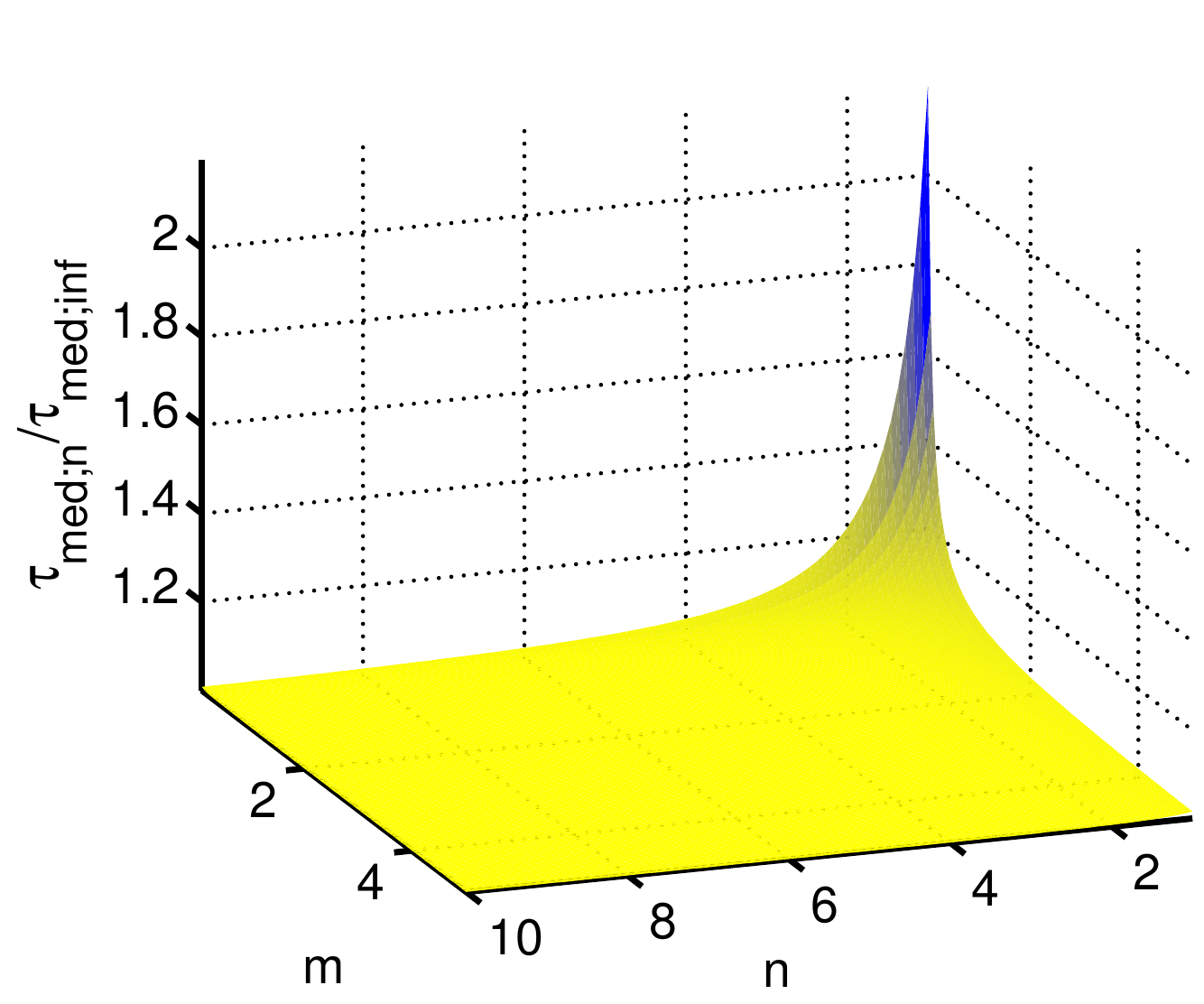}
\caption{\label{fig:tau_median_ratio_inf} Median ratio $\tau_{{\rm med};n}/\tau_{\rm med;\infty}$ for the
$\kappa$-Weibull distribution
 versus the Weibull modulus $m$ and the system size $n$. }
\end{figure}

The  median of the single RVE distribution is  defined by $R_{1}(\tau_{\rm med;1})=0.5$, and
based on~\eqref{eq:R1kappa} it is given by
$\tau_{\rm med;1}=\tau_{s} \, \left(\frac{3n}{4} \right)^{1/m}.$
The median of the $\kappa$-Weibull distribution~\cite{Clementi09} for a system of
$n = 1/\kappa$ RVEs is given by
$\tau_{{\rm med;}n} = (\ln_{\kappa}2)^{1/m}\, \tau_s$, whereas
the  median of the Weibull distribution is $ \underset{n \ra \infty} \lim  \tau_{{\rm med;}n}=(\ln 2)^{1/m}\tau_s$.
Based on the above, the ratio of the median return interval for a finite system over the median return interval of an  infinite system both of which have the same $\tau_s$, is given by
 $\tau_{{\rm med};n}/\tau_{\rm med;\infty}=\left(\ln_{\kappa}2/\ln2\right)^{1/m}$.
The ratio is plotted in Fig.~\ref{fig:tau_median_ratio_inf}.
For $n$ fixed the ratio is reduced with increasing $m$, whereas
for $m>1$ the median return interval varies only slightly with $m$.
Keeping $m$ fixed, the median return interval ratio
declines with $n$ toward 1.  This means that smaller systems have higher
  median return interval than the infinite system ---assuming that
  the characteristic interval does not change with size.
  This result is related to the heavier (i.e., power-law) upper tail of the finite-size system.

\subsection{Hazard Rate Function}

A significant question for seismic risk assessment is whether the probability
 of an earthquake of given magnitude grows or declines as the
 waiting time increases~\cite{Sornette97,Corral05a}.
 An answer to this question involves the hazard rate function of the
 return intervals.
The latter is the conditional probability that an earthquake will occur
at time  $\tau^{\ast}$ within the infinitesimal time window $\tau <\tau^{\ast} \le \tau + d\tau $,
given that there are no earthquakes in the interval $[0, \tau].$
Hence~\cite{Sornette97},
\[
h_{\tau}(\tau) = \frac{ {\rm Prob}[\tau <\tau^{\ast} \le \tau + d\tau |\tau^{\ast} > \tau ] }{d\tau} = \frac{f_{\tau}(\tau)}{R_{}(\tau)}.
\]

If earthquakes were random (memoryless) processes, distributed in time according to the Poisson law,
the ERI would follow the exponential distribution  leading to a constant $h_{\tau}(\tau)$.
If the ERI follows the Weibull distribution with cdf~\eqref{eq:Weib-cdf}, the hazard rate is given by
\beq
\label{eq:haz-weibull}
h_{\tau}(\tau)= \left( \frac{m}{\tau_s}\right)\,\left( \frac{\tau}{\tau_s}\right)^{m-1}.
\eeq
According to~\eqref{eq:haz-weibull}, the hazard rate for $m>1$ increases as $\tau \ra \infty.$
This is believed to apply to \emph{characteristic earthquakes} that occur on
faults located near plate boundaries.
In contrast, the Weibull distribution with $m<1$ as well as
the lognormal and the power-law distributions exhibit the opposite trend~\cite{Sornette97}.

 Since Bak proposed a connection between earthquakes and self-organized criticality~\cite{Bak02},
   universal or locally modified power-law expressions and the
\emph{gamma probability density function} ---which is a power law with an exponential cutoff for large times---
have been proposed  as models of the ERI pdf~\cite{Corral06a,Sornette06,Naylor09,Baro13}.
 The behavior of the gamma distribution depends on the value of the power-law exponent
  in the same way as the Weibull model.
An analysis of two earthquake catalogues based on the gamma distribution concludes that the hazard rate decreases
with time (corresponding to an exponent between 0 and 1)~\cite{Corral05a}.

The hazard rate of the $\kappa$-Weibull is given by~\eqref{eq:kw_hazard}.
For finite $n$ and for $\tau \gg \tau_s \, n^{1/m}$, $h_{\tau}(\tau) \sim 1/\tau$. If we take the
limit $n\ra \infty$ before $\tau \ra \infty$, the  Weibull hazard rate~\eqref{eq:haz-weibull} is obtained.  For a fixed
RVE size, even if $n\gg1$,  the Weibull scaling holds
for $\tau < \tau_s \, n^{1/m}$  whereas for $\tau \gg \tau_s \, n^{1/m}$
the $\tau^{-1}$ scaling dominates.
This behavior of $h(\tau)$ is demonstrated in Fig.~\ref{fig:haz_kappa}:
\begin{figure}
    \includegraphics[width=0.95\linewidth]{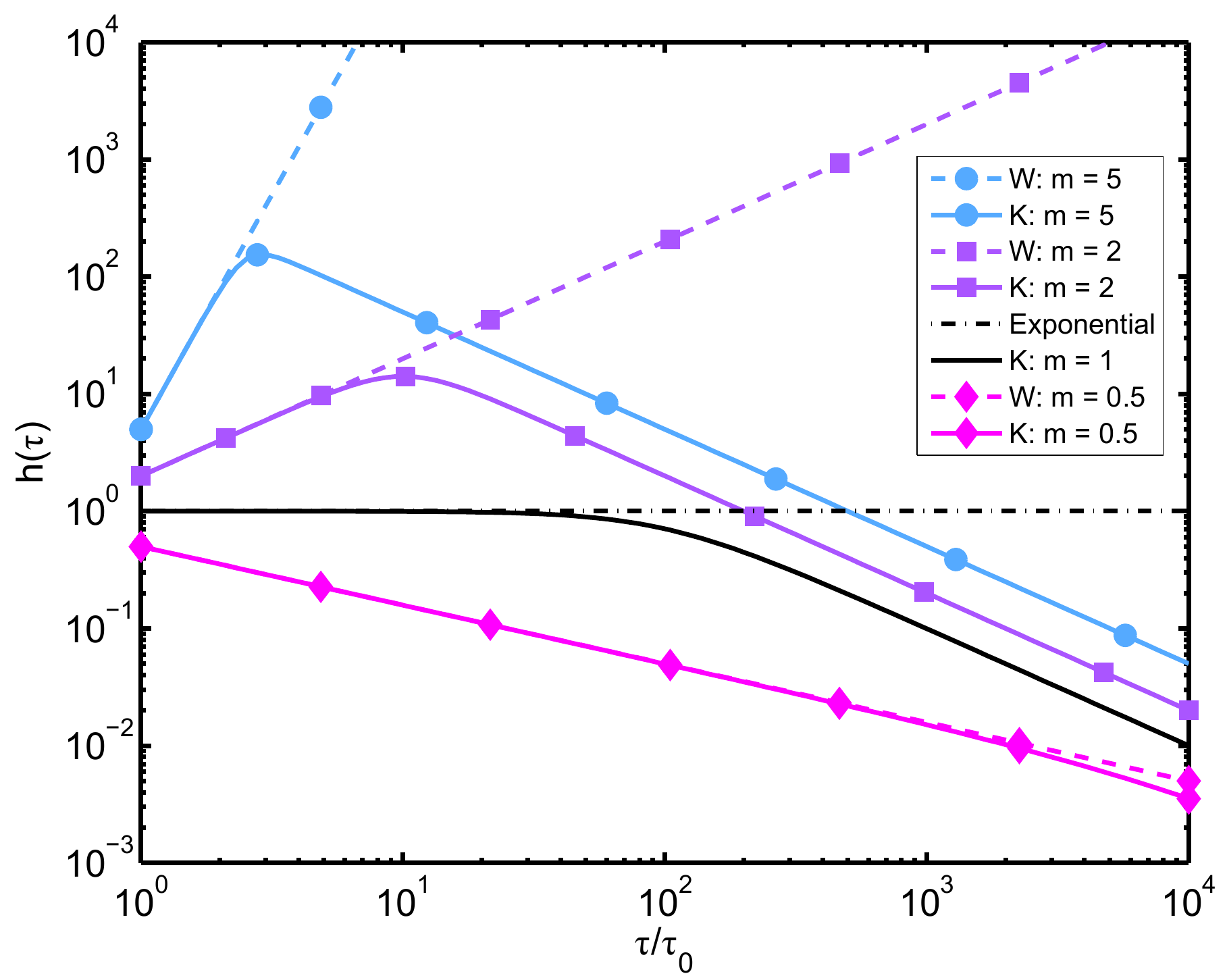}
    \caption{\label{fig:haz_kappa} Log-log plot of hazard rates $h(\tau)$ versus $\tau$ for different models. (K):  $\kappa$-Weibull
   hazard rate based on~\eqref{eq:kw_hazard} for $n=100$ with $m=0.5$ (magenta solid line and diamonds), $m=1$ (black solid line), $m=2$ (purple solid line with squares),
   and $m=5$ (cyan line with circles). (W): Weibull hazard rate obtained from~\eqref{eq:haz-weibull} with  $m=0.5$ (magenta dashed line with diamonds),
   $m=2$ (purple dashed line with squares), $m=5$ (cyan dashed line with circles).
   (Exponential): Exponential hazard rate obtained from~\eqref{eq:haz-weibull} with $m=1$ (black dash-dot line).   }
\end{figure}
For $m<1$ the dependence is not severely affected by size effects; for
$m=1$ there is a constant plateau followed by an $1/\tau$ decay, whereas for
$m>1$ the initial increase of $h(\tau)$  turns into an $1/\tau$ decay
after a turning point which occurs for $\tau_c  \approx \tau_s \, n^{1/m}.$

\section{Analysis of Earthquake Return Interval Data}
\label{sec:eri}
\begin{figure}
\centering
\includegraphics[width=0.95\linewidth]{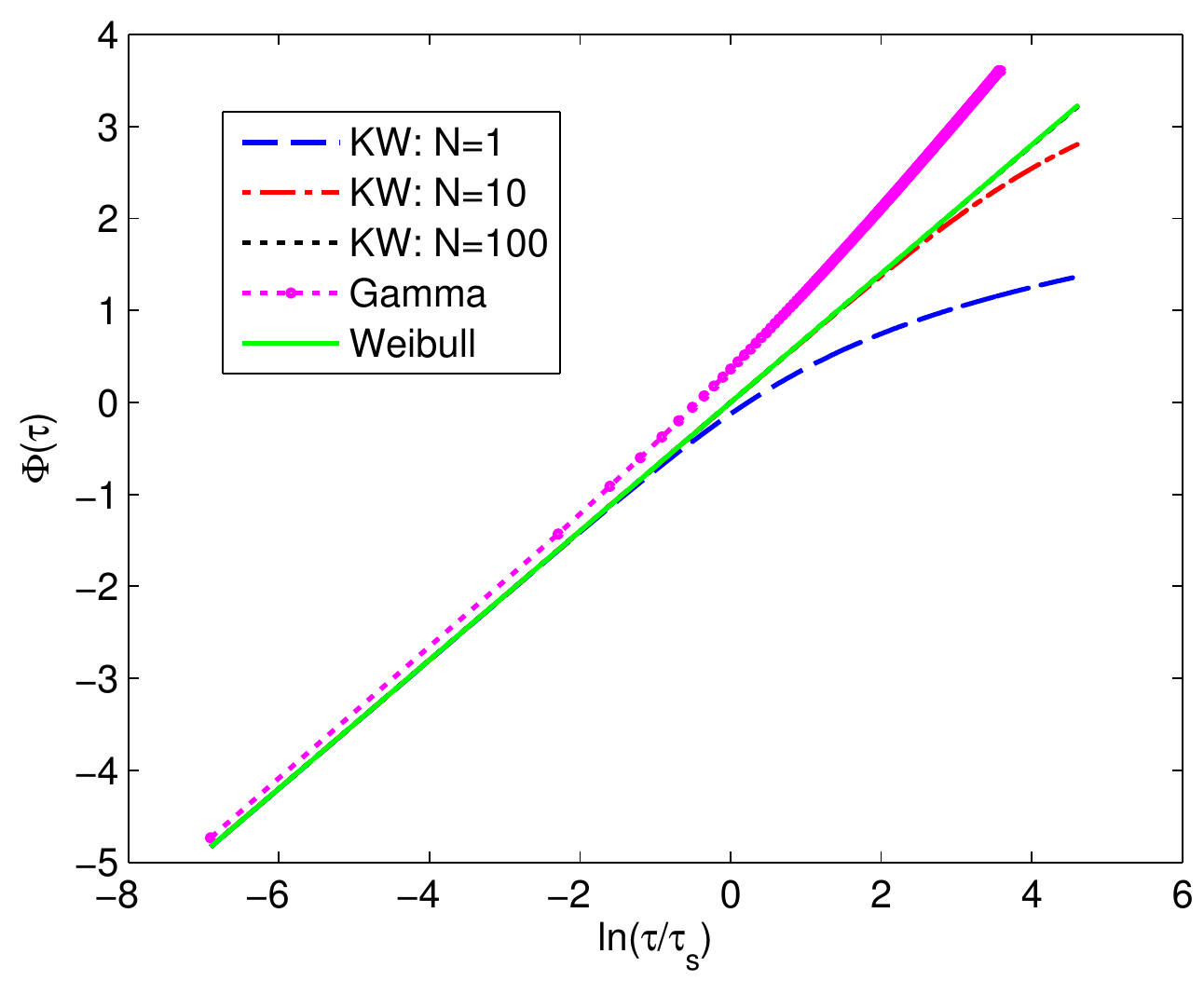}
\caption{\label{fig:Phink_Mexp_s01}
$\Phi_{n}(\tau)$ versus $\ln(\tau/\tau_s)$ for the Weibull distribution with $m=0.7$ (green solid line), the $\kappa$-Weibull with $m=0.7$ and
$n=1$ (blue dashed line), $n=10$ (red dashed and dotted line), $n=100$ (black dotted line),
and the gamma distribution with $\alpha=0.7$ (magenta dashed line with small circles). }
\end{figure}

\begin{figure}
\centering
\includegraphics[width=0.95\linewidth]{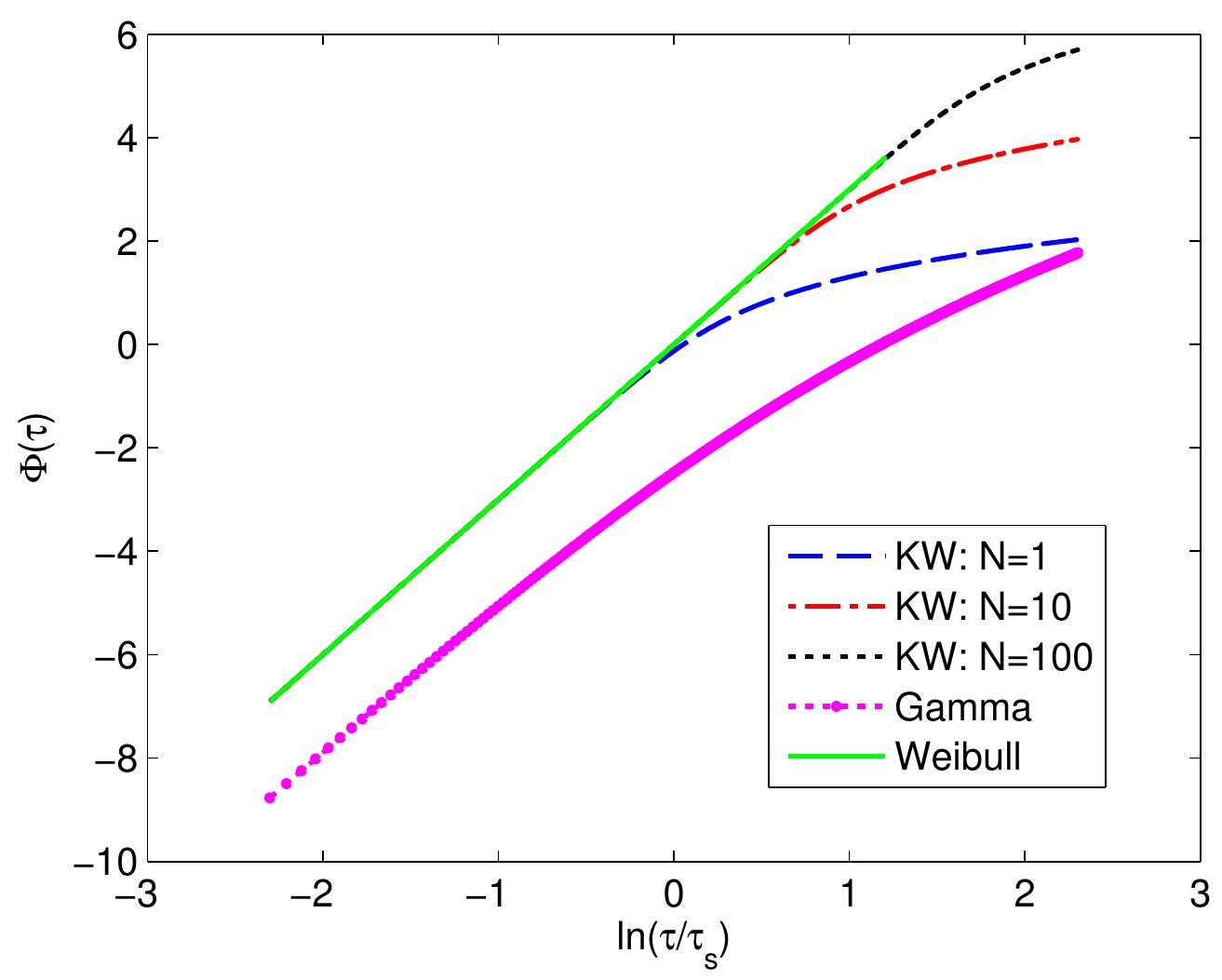}
\caption{\label{fig:Phink_Mexp_l3} $\Phi_{n}(\tau)$ versus $\ln(\tau/\tau_s)$ for the Weibull distribution with $m=3$ (green solid line),
the $\kappa$-Weibull with $m=3$ and
$n=1$ (blue dashed line), $n=10$ (red dashed and dotted line), $n=100$ (black dotted line),
and the gamma distribution with $\alpha=3$ (magenta dashed line with small circles). }
\end{figure}

The estimation of the ERI distribution from data is complicated by the fact that
the $\kappa$-Weibull distribution
and the Weibull distribution are close over the range  $0 \le \tau \le \tau_w$,
where $\tau_w$ is a parameter
that depends on $n$ and $m$.
Differences in the tail of ERI distributions are best visualized in terms of  $\Phi_{n}(\tau)$, as shown in
the Weibull plots of Figs.~\ref{fig:Phink_Mexp_s01}-\ref{fig:Phink_Mexp_l3}.
On these diagrams, the deviation of the $\kappa$-Weibull distribution
from the straight line diminishes with increasing $n$.
The gamma distribution is also included
for comparison purposes. The gamma probability model with pdf $f(\tau)= \tau^{\alpha-1}\exp(-\tau/b)\big/ b^{\alpha}\Gamma(\alpha)$ is often used in studies of
 earthquake return intervals, e.g.~\cite{Corral03},\cite{Corral05a}.
 For $m<1$ the Weibull plot of the gamma probability distribution
 is a convex function, whereas for $m>1$
it becomes concave. In contrast, the $\kappa$-Weibull distribution is concave for all $m$.

\subsection{Microseismic sequence from Crete}
We consider the return intervals for an earthquake sequence from the island of Crete (Greece)
 which involves over 1\,821 micro-earthquake events with magnitudes up to
 4.5 (${\rm M_L}$)~(Richter local magnitude scale)~\cite{dth12}.
 The sequence was recorded between July 2003 and June 2004~\cite{dth12}.
The return intervals between successive earthquake events range from 1 (sec)
to 19.5 (days). The spatial domain covered is approximately between $24.5^\circ$ -- $27^\circ$ (East longitude) and
$34^\circ$ -- $35.5^\circ$ (North latitude). The magnitude of completeness for this data set is around
2.2 -- 2.3 (${\rm M_L}$), which means that all events exceeding this magnitude are registered by the
measurement network.

\begin{figure}
\includegraphics[width=0.75\linewidth]{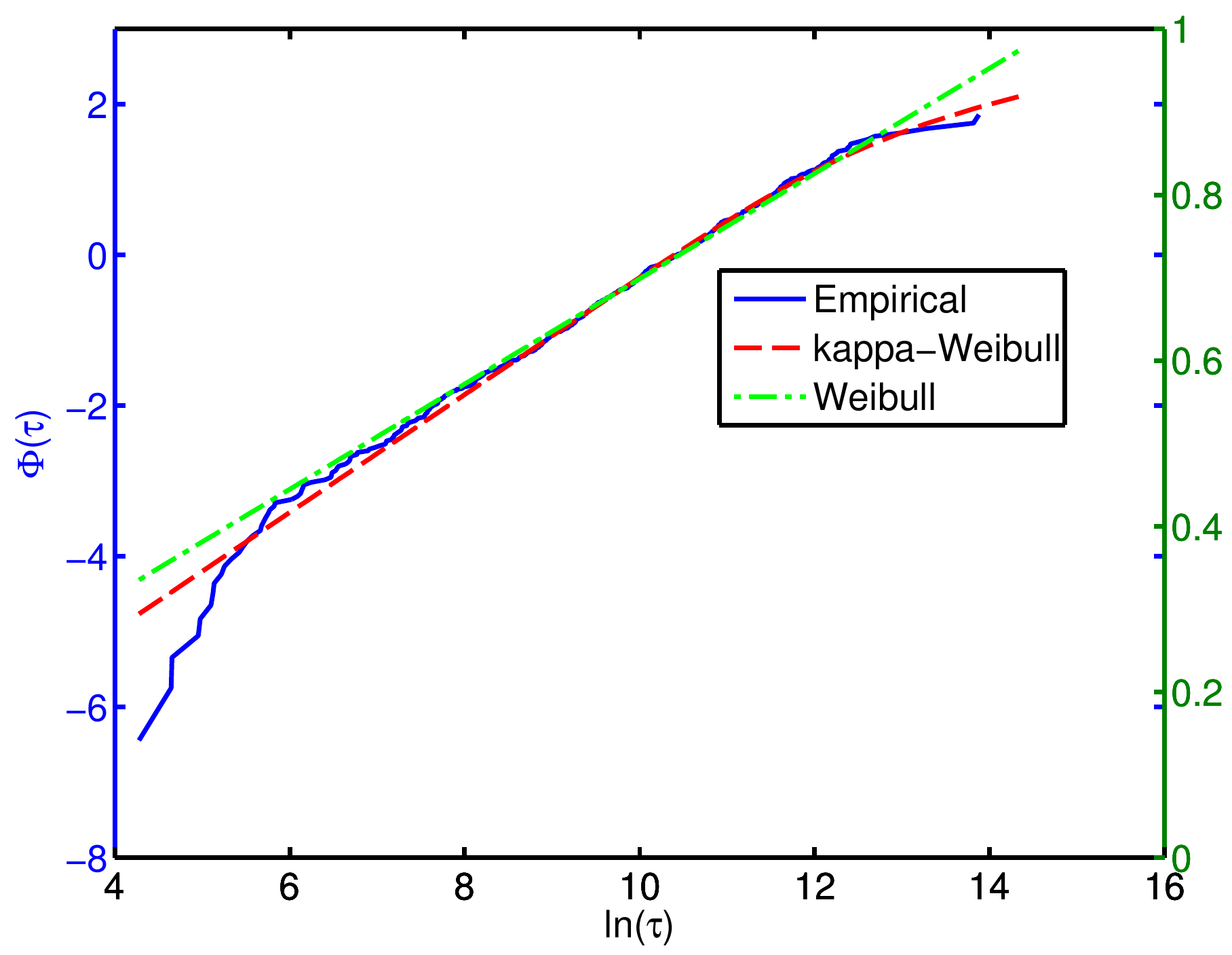}
\caption{\label{fig:Crete_phi_wblk_2p3}
$\Phi(\tau)$ versus $\ln(\tau)$ for  earthquake return intervals of
 the Cretan earthquake sequence (CES). The return intervals refer to 628 events with magnitudes exceeding
2.3  (${\rm M_L}$).  The magnitude of completeness  is $\approx 2.2 $ (${\rm M_L}$).
The  maximum likelihood estimates of the $\kappa$-Weibull parameters are
$\hat{\tau}_{0} \approx 3.19\times 10^4$ (sec),  $\hat{m} \approx 0.78$,
$\hat{\kappa} \approx 0.33$. The estimate of $\Phi(\tau)$ using the empirical (data-based) cdf is shown with the solid blue line. The
optimal $\kappa$-Weibull fit is shown with the red dashed line, whereas the optimal Weibull fit is shown with the green dashed and dotted line. The left vertical axis measures $\Phi(\tau)$, whereas the right vertical axis marks
the corresponding cdf values. }
\end{figure}

\begin{figure}
\includegraphics[width=0.75\linewidth]{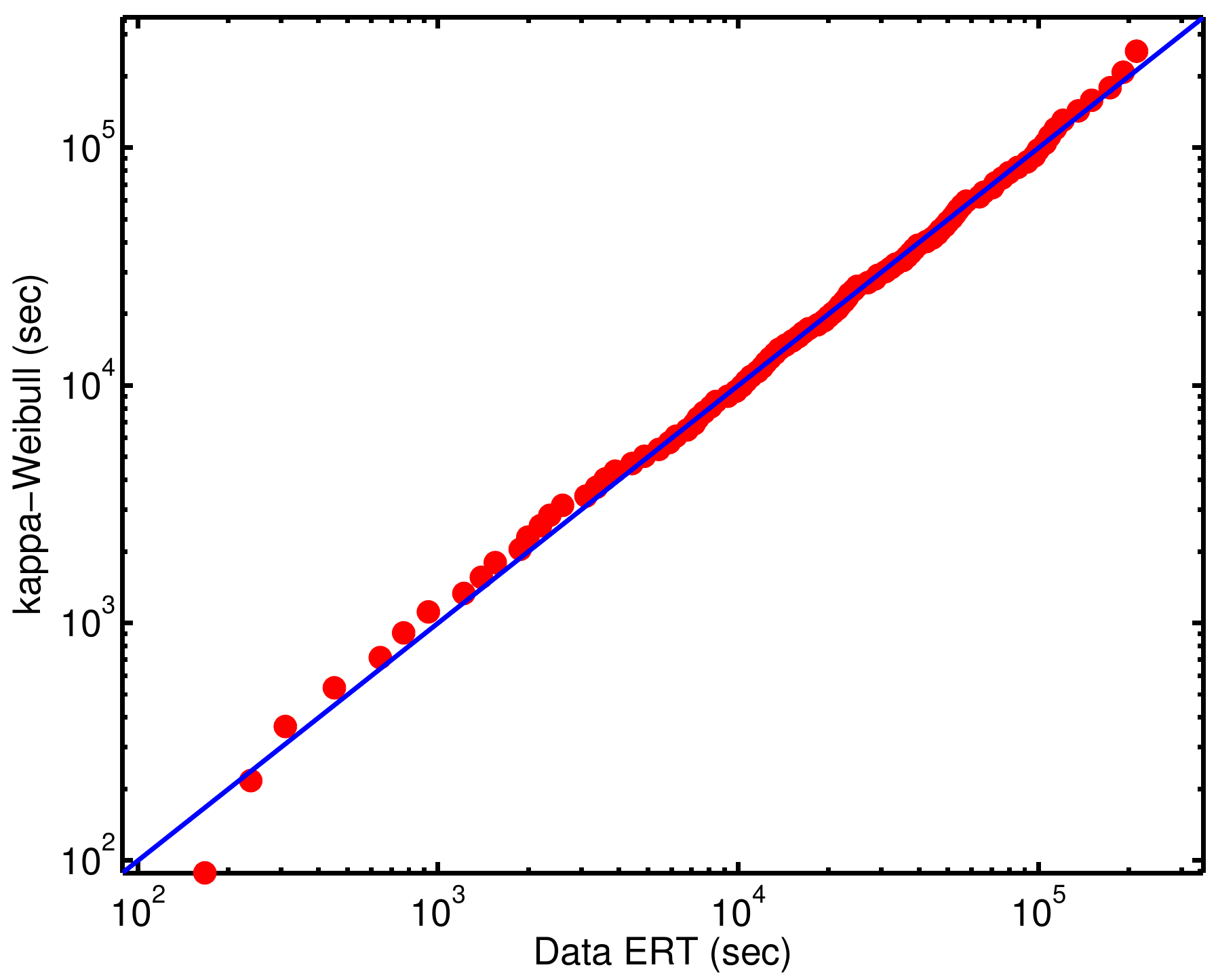}
\caption{\label{fig:Crete_qq_wblk_2p3}
Sample quantiles   versus the corresponding quantiles of the optimal $\kappa$-Weibull distribution for
 the  Cretan earthquake  sequence return intervals (CES ERI) ---$\Phi(\tau)$ of the data shown in Fig.~\ref{fig:Crete_phi_wblk_2p3}.
 The $\kappa$-Weibull quantiles are obtained from~\eqref{eq:kw_quantile} using the maximum likelihood estimates
 of the $\kappa$-Weibull parameters (see caption of Fig.~\ref{fig:Crete_phi_wblk_2p3}).}
\end{figure}

We use the  method of \emph{maximum likelihood} to estimate
the parameters of test probability distributions for
the return intervals.
The optimal $\kappa$-Weibull distribution
for events above $M_{c}=2.3$ (${\rm M_L}$) is compared with the optimal Weibull distribution in Fig.~\ref{fig:Crete_phi_wblk_2p3}~\footnote{We use the MATLAB
{\it fmincon} constrained minimization function with a trust region reflective algorithm and explicit
gradient information to minimize the negative log-likelihood. The method is applied iteratively (i.e., {\it fmincon} is called with the
last estimate as initial guess) for fifty times. The optimization parameters used are: Maximum number of objective
function evaluation $=2\times 10^4$, maximum number of iterations $=2\times 10^4$, objective function error tolerance $=1\times 10^{-5}$.
Tests with the data sequences showed that this number of iterations is sufficient for
the parameter estimates to converge.
The accuracy and precision of the estimates  were also tested using synthetic sequences of $\kappa$-Weibull random numbers
generated by means of the inversion method. }.
Note that the empirical distribution of the return intervals has $m<1$ and a concave tail,
in contrast with the gamma density model (cf. Fig.~\ref{fig:Phink_Mexp_s01}).
The $\kappa$-Weibull distribution approximates better the upper tail of the return intervals than the Weibull distribution.
 Both the Weibull distribution and the $\kappa$-Weibull distribution have a lighter lower tail than the data.
These trends persist as $M_{c} \uparrow$, but the differences between the distributions progressively decrease.
Fig.~\ref{fig:Crete_qq_wblk_2p3} compares the quantiles of the data distribution
with those of the optimal $\kappa$-Weibull model.


We investigate different hypotheses for the ERI distribution
using the \emph{Kolmogorov-Smirnov}  test following the methodology described in~\cite{Clauset09}.
The Kolmogorov-Smirnov distance between the empirical (data) distribution, ${F}_{\rm emp}(\tau)$ and the estimated (model)
distribution, $\hat{F}(\tau)$, is given by $D = \sup_{\tau \in \mathbb{R}} | {F}_{\rm emp}(\tau) - \hat{F}(\tau) |$, where
$\sup_{A}f(\tau)$ denotes the supremum of $f(\tau)$ for $ \tau \in A$.
The parameters of $\hat{F}(\tau)$ are also estimated using the method of maximum likelihood as described above.
The null hypothesis is that $\hat{F}(\tau)$  represents the probability distribution of the data.
We apply the test to the Poisson, normal (Gaussian), lognormal, Weibull, $\kappa$-Weibull,  gamma, and generalized
gamma distributions. The generalized gamma distribution~\cite{stacy62}, with pdf given by
$f(x) = (d/x_{s})^{m} \, x^{d-1} \, {\rm e}^{-(x/x_{s})^m} \big/ \Gamma(d/m)$, incorporates both the
gamma distribution (for $m=1$) and the Weibull distribution (for $m=d$).

The Kolmogorov-Smirnov test for a probability model with estimated parameters should be applied using
Monte Carlo simulation to generate synthetic data from the estimated probability  model.
We generate  random numbers from the Poisson, normal, lognormal, Weibull, and gamma distributions
with the respective MATLAB random number generators. For the $\kappa$-Weibull ($\kappa>0$) and for
the generalized gamma distribution we implemented
the inverse transform sampling method (see Appendix~\ref{App:A}).
For each realization of return intervals, we estimate the parameters of the optimal distribution model, $\hat{F}^{(j)}(\tau), j=1,\ldots, N_{\rm sim}$. The
 Kolmogorov-Smirnov distance between the empirical distribution of the specific realization, ${F}^{(j)}_{\rm emp}(\tau)$,
 and the estimated model distributions
 for the $j^{\rm th}$ realization are given by
$D^{(j)} = \sup_{\tau \in \mathbb{R}} | {F}^{(j)}_{\rm emp}(\tau) - \hat{F}^{(j)}(\tau) |,
\; j= 1, \ldots,N_{\rm sim}$.
The $p$-value of the Kolmogorov-Smirnov distance is defined as
$p= \frac{1}{N_{\rm sim}} \sum_{i=j}^{N_{\rm sim}}\mathbbm{1}_{D^{(j)} > D}(D^{(j)})$
---where $\mathbbm{1}_{A}(\tau)=1$, if $\tau \in A$ and $\mathbbm{1}_{A}(\tau)=0$, if $\tau \notin A$, is
the indicator function of the set $A$. The $p$-value is the probability  the Kolmogorov-Smirnov distance
will exceed $D$ purely by chance if the null hypothesis is true. If $p > p_{\rm crit}$,  the null hypothesis is accepted, otherwise it is rejected.

If we focus on the return intervals between earthquakes with magnitudes exceeding $2.3$,
the Kolmogorov-Smirnov test (based on  $1\,000$ simulations) rejects the normal, gamma, and lognormal and Poisson distributions at the $5\%$ significance level.
 In contrast, the  Weibull and $\kappa$-Weibull distributions are accepted with $p \approx 0.09$
and $p \approx0.75$, respectively, whereas the generalized gamma at $p \approx 0.10$.
For $3.5 \ge M_c \ge 2.3$ (${\rm M_L}$)  the calculated $p$-values are
shown in Table~\ref{t:Crete-P}. The following remarks summarize the  tabulated results:
(i) The $\kappa$-Weibull and the Weibull distributions are accepted for all $M_c$ except for 2.7 and 3.5.
(ii) The gamma distribution is accepted for $M_c \ge 2.5$  whereas the generalized gamma is accepted for all
$M_c$ except for $M_c =2.7$.
(iii) The lognormal is marginally accepted for $M_c = 3.5$.
(iv) The Poisson distribution is accepted for $M_c = 3.7, 3.9$
(v) The normal and Poisson models are rejected for all $M_c \le 3.7$.
(vi) For $M_c \le 3.3$ the Weibull, the gamma, generalized gamma and the $\kappa$-Weibull models have the  highest $p$-values but their relative
ranking changes with $M_c$. Note that for $M_c > 3.3$ the sample size is quite small $(45 \ge N_c \ge 18 )$ thus prohibiting
the observation of long tails.
\begin{table}
\begin{ruledtabular}
\caption{$p$-values of Kolmogorov-Smirnov test for the fit between different probability models and the Crete Earthquake Sequence.
Results are based on $1\,000$ simulations. The
sample size (number of return intervals) is shown within parentheses next to the cutoff magnitudes.}
\label{t:Crete-P}
\begin{tabular}{lccccccc}
&\textbf{Gamma}&\textbf{Weibull}&\textbf{$\kappa$-Weibull}&\textbf{Gen. Gamma}&\textbf{Normal}&\textbf{Lognormal}&\textbf{Poisson}\\
{$M_{L,c}=$2.3 (628)}&0.00&0.09&0.75&0.10&0.00&0.00&0.00\\
{$M_{L,c}=$2.5 (414)}&0.07&0.40&0.14&0.35&0.00&0.00&0.00\\
{$M_{L,c}=$2.7 (273)}&0.38&0.01&0.00&0.04&0.00&0.00&0.00\\
{$M_{L,c}=$2.9 (176)}&0.14&0.23&0.20&0.25&0.00&0.00&0.00\\
{$M_{L,c}=$3.1 (103)}&0.28&0.27&0.20&0.40&0.00&0.00&0.00\\
{$M_{L,c}=$3.3 (69)}&0.28&0.11&0.10&0.19&0.00&0.01&0.00\\
{$M_{L,c}=$3.5 (45)}&0.15&0.03&0.03&0.08&0.00&0.05&0.00\\
{$M_{L,c}=$3.7 (28)}&0.16&0.10&0.08&0.07&0.00&0.00&0.14\\
{$M_{L,c}=$3.9 (18)}&0.20&0.19&0.17&0.09&0.27&0.02&0.25\\
\end{tabular}
\end{ruledtabular}
\end{table}

Since for most $M_c$ more than one model hypotheses pass the Kolmogorov-Smirnov test, it is desirable to  somehow
compare the different probability models. For this purpose we use the Akaike Information Criterion (AIC)~\cite{Akaike74}.
The AIC is  defined by
$\mathrm{AIC} = 2 \,\mathrm{NLL} + 2 k$, where $\mathrm{NLL}$ is the negative log-likelihood
of the data for the given model, and
$k$ is the number of model parameters ($k=1$ for the Poisson, $k=2$ for the normal,
lognormal, Weibull, and gamma, whereas $k=3$ for the $\kappa$-Weibull and the generalized gamma).
The term $2k$ in AIC penalizes models with more parameters.
In general, a model with
lower AIC is preferable to one with higher AIC.
We present AIC results for the Crete Earthquake Sequence in Table~\ref{t:Crete-AIC}. The tabulated values
correspond to $\mathrm{AIC} /N_c$.
The following conclusions can be reached from this Table: (i) The
gamma, generalized Weibull, and $\kappa$-Weibull distributions have similar
AIC values which are lower  than the normal, lognormal, and Poisson models.
(ii) The AIC values of the four top ranking distributions are quite close to each other.
(iii) The $\kappa$-Weibull  has the lowest AIC for the
larger samples (i.e., those with $M_c =2.3, 2.5$).
The observation (i) also explains the somewhat unexpected outcome of Table~\ref{t:Crete-P},
namely, that the $p$-values of the generalized gamma and the $\kappa$-Weibull are not ---for all
magnitude cutoffs--- equal or higher
than  the $p$-values  of the respective subordinated distributions, i.e., the gamma  and the Weibull respectively:
The estimates of the probability model parameters are based on the minimization
of the negative log-likelihood, which provides a different
measure of the fit between the data and the model distribution than the Kolmogorov-Smirnov distance.
We checked that the
 incongruence  remains even if the likelihood optimization algorithm for the generalized gamma and the $\kappa$-Weibull is initialized
by the respective optimal parameters of the gamma and Weibull distributions for the same data set.

\begin{table}
\begin{ruledtabular}
\caption{Akaike Information Criterion (AIC) values per sample point for different probability models. The model parameters
used maximize the likelihood  of the Crete Earthquake Sequence given the respective model.
The sample size (number of return intervals) is shown within parentheses next to the cutoff magnitudes.}
\label{t:Crete-AIC}
\begin{tabular}{lccccccc}
&\textbf{Gamma}&\textbf{Weibull}&\textbf{$\kappa$-Weibull}&\textbf{Gen.~Gamma}&\textbf{Normal}&\textbf{Lognormal}&\textbf{Poisson}\\
{$M_{L,c}=$2.3 (628)}&23.23&23.16&23.12&23.15&25.95&23.21&23.46\\
{$M_{L,c}=$2.5 (414)}&24.05&24.02&24.00&24.02&26.40&24.12&24.29\\
{$M_{L,c}=$2.7 (273)}&24.91&24.90&24.90&24.90&26.89&25.07&25.12\\
{$M_{L,c}=$2.9 (176)}&25.79&25.78&25.79&25.79&27.59&25.91&26.00\\
{$M_{L,c}=$3.1 (103)}&26.81&26.80&26.82&26.82&28.56&26.92&27.08\\
{$M_{L,c}=$3.3 (69)}&27.58&27.59&27.62&27.62&29.23&27.76&27.86\\
{$M_{L,c}=$3.5 (45)}&28.17&28.20&28.24&28.23&30.22&28.38&28.57\\
{$M_{L,c}=$3.7 (28)}&29.30&29.36&29.43&29.41&30.67&29.79&29.46\\
{$M_{L,c}=$3.9 (18)}&30.37&30.43&30.54&30.50&31.08&30.92&30.38\\
\end{tabular}
\end{ruledtabular}
\end{table}

\subsection{Southern California Data}
We also analyze an earthquake sequence which contains 2\,446  events  in Southern California
($114^\circ$ -- $122^\circ$ West longitude and $32^\circ$ -- $37^\circ$ North latitude) down to depths of $\le 20$ (km)
 with magnitudes from 1 (except for 3 events at 0.5) up to  6.5 (${\rm M_L}$); 2\,444 of these events have magnitudes less than 5.0 (${\rm M_L}$),
 whereas the two main shocks have 6.0 (${\rm M_L}$) and 6.5 (${\rm M_L}$). The events occurred during the period from
 January 1, 2000 until March 27, 2012~\footnote{The facilities of the Southern Californian earthquake Data Center (SCEDC),
 and the Southern California Seismic Network (SCSN), were used for access to waveforms, parametric data,
 and metadata required in this study.  The SCEDC and SCSN are funded through U.S. Geological Survey Grant G10AP00091,
 and the Southern Californian earthquake Center, which is funded by NSF Cooperative Agreement EAR-0529922 and USGS Cooperative Agreement 07HQAG0008.}.

The $p$-values of the Kolmogorov-Smirnov test are listed in Table~\ref{t:SC-P}.
The Weibull and $\kappa$-Weibull distributions have practically the same
$p$-values.
The gamma and generalized gamma models, however, show overall better agreement with the observed return intervals than
the Weibull or the $\kappa$-Weibull.  Most of the $p$-values obtained for  this data set are considerably lower than
their counterparts for the Cretan data set. To ensure that this difference is not caused by an insufficient number of Monte Carlo
simulations, we repeated the numerical experiment with $5\,000$ Monte Carlo simulations, which confirmed the
results of Table~\ref{t:SC-P} with minor changes in the $p$-values.  On the other hand,
as shown in Table~\ref{t:SC-AIC},  the gamma, generalized gamma, Weibull and $\kappa$-Weibull distributions
 have similar AIC values. The low $p$-values are an indication that none of the models tested match the data
 very well in terms of the Kolomogorov-Smirnov distance. The gamma, generalized gamma, Weibull and $\kappa$-Weibull, 
 however, are not rejected at the $1\%$ level for most thresholds. It should be noted that 
 recent arguments based on Bayesian analysis of hypothesis testing
 suggest that the significance level $0.05$ used to  reject the null hypothesis is overly
 conservative and should be shifter to $0.005$~\cite{Johnson13}. 

\begin{table}
\begin{ruledtabular}
\caption{$p$-values of Kolmogorov-Smirnov test for the fit between different probability models and the Southern California sequence.
Results are based on $5\,000$ simulations.  The
sample size (number of return intervals) is shown within parentheses next to the cutoff magnitudes.}
\label{t:SC-P}
\begin{tabular}{lccccccc}
&\textbf{Gamma}&\textbf{Weibull}&\textbf{$\kappa$-Weibull}&\textbf{Gen.~Gamma}&\textbf{Normal}&\textbf{Lognormal}&\textbf{Poisson}\\
{$M_{L,c}=$2.3 (1341)}&0.0296&0.0000&0.0000&0.0000&0.0000&0.0000&0.0000\\
{$M_{L,c}=$2.5 (964)}&0.0368&0.0000&0.0000&0.0000&0.0000&0.0000&0.0000\\
{$M_{L,c}=$2.7 (687)}&0.0348&0.0000&0.0000&0.0000&0.0000&0.0000&0.0000\\
{$M_{L,c}=$2.9 (457)}&0.0908&0.0002&0.0004&0.0000&0.0000&0.0000&0.0000\\
{$M_{L,c}=$3.1 (309)}&0.5036&0.0638&0.0556&0.0440&0.0000&0.0000&0.0000\\
{$M_{L,c}=$3.3 (206)}&0.0254&0.0158&0.0150&0.0270&0.0000&0.0000&0.0000\\
{$M_{L,c}=$3.5 (121)}&0.1836&0.0110&0.0086&0.0152&0.0000&0.0000&0.0000\\
{$M_{L,c}=$3.7 (77)}&0.0798&0.0304&0.0258&0.0412&0.0000&0.0002&0.0000\\
{$M_{L,c}=$3.9 (44)}&0.0150&0.0450&0.0378&0.0580&0.0000&0.0964&0.0000\\
\end{tabular}
\end{ruledtabular}
\end{table}

\begin{table}
\begin{ruledtabular}
\caption{Akaike Information Criterion (AIC) values per sample point for the best-fit models of the Southern California sequence.
The model parameters  used maximize the likelihood  of the Southern California sequence given the respective model.
The sample size (number of return intervals) is shown within parentheses next to the cutoff magnitudes.}
\label{t:SC-AIC}
\begin{tabular}{lccccccc}
&\textbf{Gamma}&\textbf{Weibull}&\textbf{$\kappa$-Weibull}&\textbf{Gen.~Gamma}&\textbf{Normal}&\textbf{Lognormal}&\textbf{Poisson}\\
{$M_{L,c}=$2.3 (1341)}&26.45&26.45&26.45&26.44&29.19&26.68&27.14\\
{$M_{L,c}=$2.5 (964)}&26.98&27.01&27.01&27.01&29.73&27.26&27.79\\
{$M_{L,c}=$2.7 (687)}&27.55&27.59&27.59&27.60&30.33&27.85&28.47\\
{$M_{L,c}=$2.9 (457)}&28.36&28.41&28.42&28.40&30.92&28.65&29.29\\
{$M_{L,c}=$3.1 (309)}&28.99&29.02&29.03&29.02&31.82&29.22&30.06\\
{$M_{L,c}=$3.3 (206)}&29.28&29.30&29.31&29.31&32.69&29.43&30.88\\
{$M_{L,c}=$3.5 (121)}&30.03&30.07&30.08&30.08&33.84&30.22&31.84\\
{$M_{L,c}=$3.7 (77)}&30.64&30.72&30.81&30.72&34.65&30.89&32.74\\
{$M_{L,c}=$3.9 (44)}&30.70&30.67&30.72&30.72&35.68&30.67&33.46\\
\end{tabular}
\end{ruledtabular}
\end{table}

\section{Fiber Bundle Models}
\label{sec:fbm}
Fiber Bundle Models (FBM) are simple statistical models that were introduced  to study the fracture of fibrous materials~\cite{Daniels45}.
To date they are used in many research fields, including  fracture of composite materials~\cite{Phoenix1983}, landslides~\cite{Cohen2009},
glacier avalanches~\cite{Reiweger2009} and earthquake dynamics~\cite{Chakrabarti97,Alava06,Abaimov08}.
In spite of their conceptual simplicity, FBMs exhibit surprisingly rich behavior.

\begin{figure}
    \centering
    \includegraphics[width = 0.33 \linewidth]{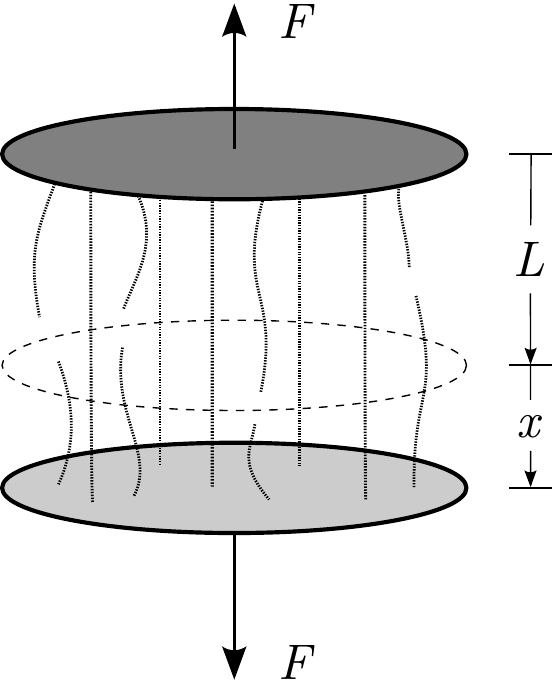} 
    \caption{Schematic of a fiber bundle of initial length $L$ elongated by $x$ due to the loading force $F$.
    Broken fibers do not contribute to the strength of the bundle.}
    \label{fig:schematic}
\end{figure}

An FBM consists of an arrangement of parallel fibers subject to an external load $F$ (Fig.~\ref{fig:schematic}).
The fibers have random strength thresholds that represent the heterogeneity of the medium.
Due to the applied loading, each fiber is deformed and subject to stress.
If the stress applied to a specific fibre exceeds its failure threshold, the fiber ruptures and the excess load is redistributed
either globally or locally between the remaining fibers. The ensuing redistribution of the load to the surviving fibers may trigger an avalanche of breaks.
Each fibre break releases  the elastic energy accumulated in the fibre.

\subsection{FBM Return Interval Statistics}

\begin{figure}
    \centering
    \includegraphics[width = 0.75\linewidth, trim = 20 0 20 0]{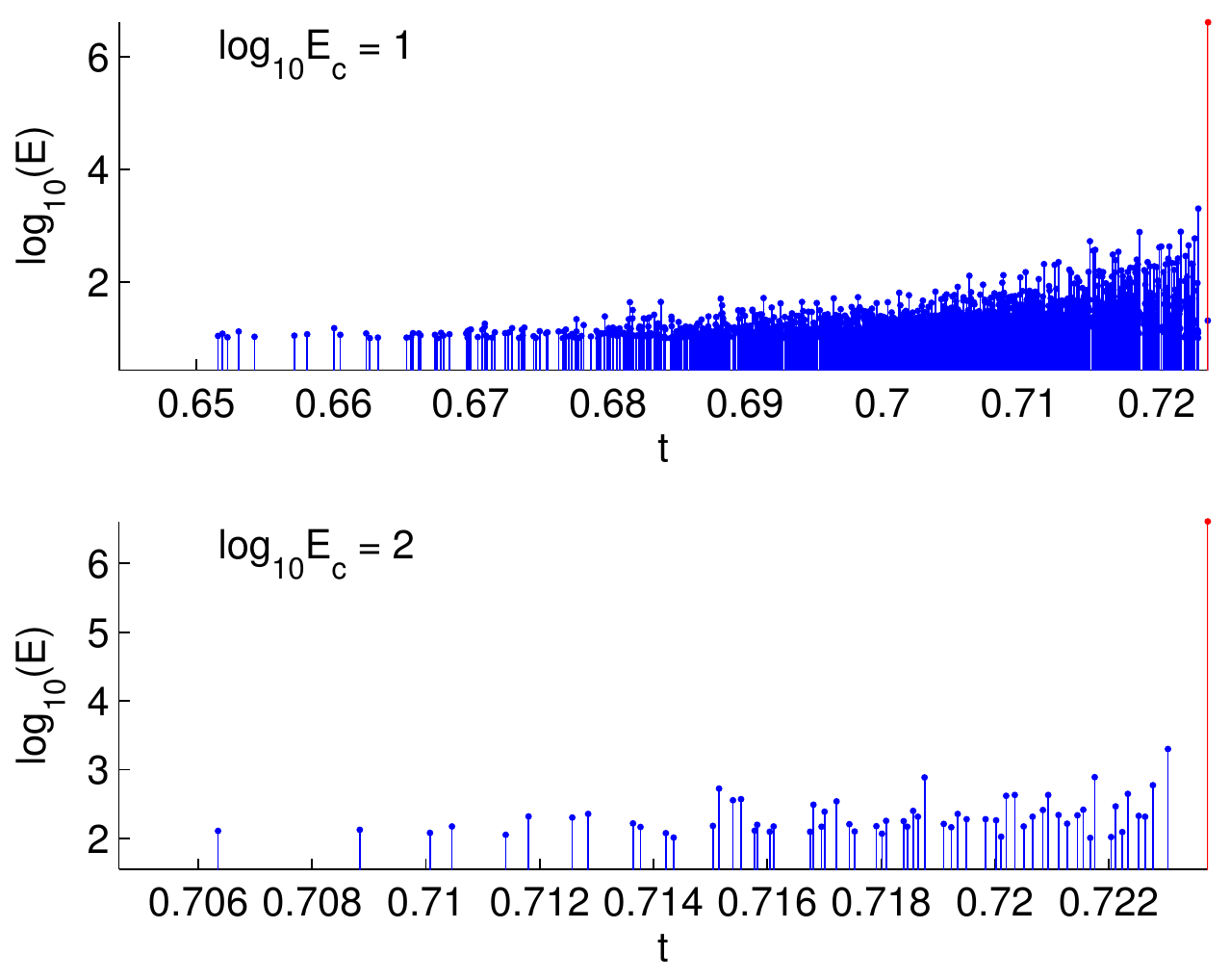}
    \caption{Evolution  of avalanche events in time for a bundle of $10^7$ fibers with
     Weibull-distributed strength thresholds ($m=5, x_s=1$).
    The most energetic avalanche $(\log_{10} E_f \approx 7.31)$  is generated when the bundle fails (during the last event of the breaking sequence) at time $t_f \approx 0.725$.}
    \label{fig:timelines}
\end{figure}

We assume that the strain $\epsilon$ of the fiber bundle increases linearly
 with time $t$, i.e., $\epsilon \propto t$.
Without loss of generality we set the elastic modulus, the initial length $L$ and the strain rate equal to unity,
and we  use the elongation $x$
instead of $\epsilon$ to measure the loading.
The individual fibers have random failure thresholds $x_{\rm c}$ with pdf $\fx(x)$.
Failed fibers  are removed,
and the stress is then redistributed between the surviving fibers using the equal load sharing rule.
The energy of each avalanche is equal to the sum of the Hookean energies of the broken fibers~\cite{Pradhan2008}.
Only events that exceed an energy threshold $E_{\rm c}$ are counted. The return intervals are measured as the time difference
between two events with energy $E \ge E_{\rm c}$. Avalanches are considered to occur instantaneously.
Fig.~\ref{fig:timelines} illustrates the evolution  of the avalanche events in time. The plots are
obtained by loading a single bundle of $10^7$ fibers the strength of which follows the Weibull pdf with $m=5, x_s =1$.
The avalanche sequences correspond to energy thresholds given by  $\log_{10} E_c = 1, 2$, where $\log_{10}$ is the
logarithm with base 10.

\begin{figure}
    \centering 
    \includegraphics[width = 0.8\linewidth]{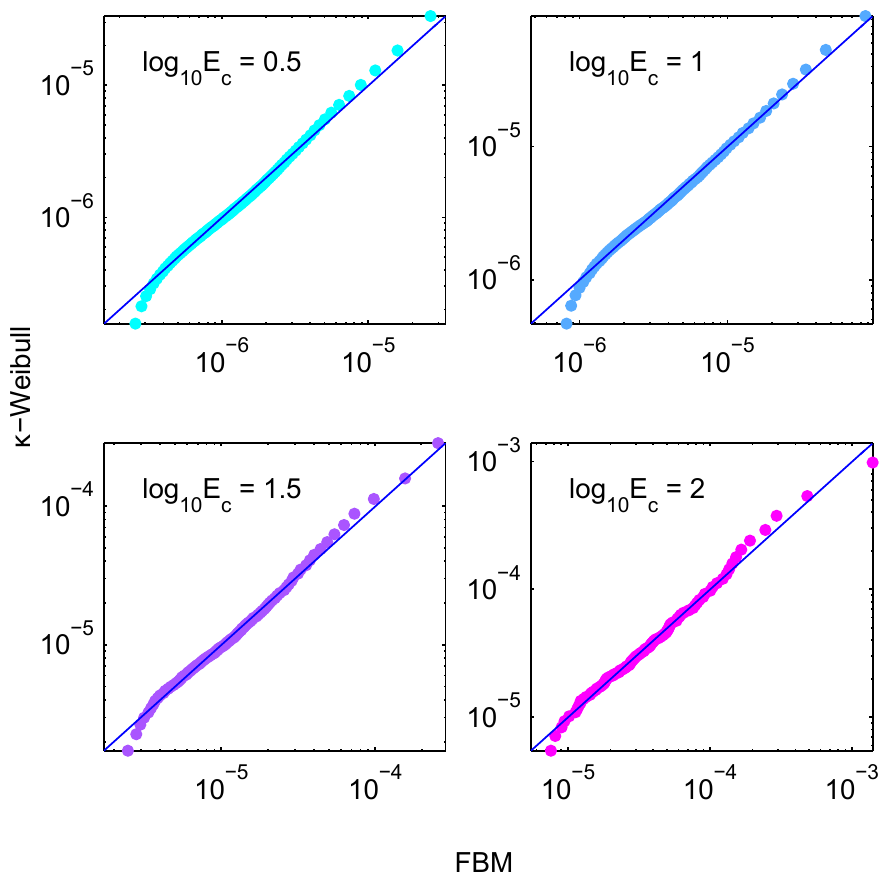}
    \caption{Quantile-quantile  plots of FBM return intervals (dimensionless) versus the best-fit $\kappa$-Weibull distribution.
    Data are based on a bundle of $5 \times 10^7$ fibers with strength thresholds drawn from the
    Weibull distribution with $m=5$, $x_s=1$, unit elastic constant, and unit strain rate.}
    \label{fig:fbm-QQ}
\end{figure}

Figure~\ref{fig:fbm-QQ} compares the quantiles of the return interval distribution obtained from a single bundle
with those of the optimal $\kappa$-Weibull distribution for different energy thresholds. The optimal $\kappa$-Weibull
parameters for each threshold are shown in Table~\ref{t:1}.
The estimated $\kappa$ values based on maximum likelihood are $\hat{\kappa} \approx 2$.
This result suggests that $\kappa >1$ values indicate a highly correlated system that can not
be decomposed into RVEs.
As stated in Subsection~\ref{subsec:kweibull}, the $\kappa$-Weibull pdf exhibits a power-law upper tail with exponent $\alpha = 1 + m/\kappa$.
For the FBM investigated above, the exponent of the return interval pdf is $\hat{\alpha}  \approx [2.13, 2.20]$.

\begin{table}
\begin{ruledtabular}
    \caption{Maximum Likelihood estimates of $\kappa$-Weibull distribution parameters for the return intervals of Figure~\ref{fig:fbm-QQ}.}
    \label{t:1}
    \begin{tabular}{c c c c c c c c c c}
        $\log_{10} E_c$ & $N$            & $\hat{\tau}_s$       & $\hat{m}$ & $\hat{\kappa}$ \\ \hline
        0.5        & 44094          & $1.2 \times 10^{-6}$ & 2.4       & 2.1            \\
        1          & 8449           & $3.7 \times 10^{-6}$ & 2.4       & 2.0            \\
        1.5        & 1686           & $1.0 \times 10^{-5}$ & 2.6       & 2.2            \\
        2          & 311            & $3.1 \times 10^{-5}$ & 2.6       & 2.3            \\
    \end{tabular}
\end{ruledtabular}
\end{table}


\section{Discussion and Conclusions}
\label{sec:concl}
We investigated the statistics of return intervals in systems that obey weakest-link scaling.  We propose that the $\kappa$-Weibull distribution is suitable for finite-size systems (where the size is
measured in terms of RVE size) and that the parameter $\kappa$ is determined by the size of the system.
A characteristic property of the $\kappa$-Weibull distribution is the
transition from Weibull to power-law scaling in the upper tail of the pdf.
This leads to an upper tail which decays slower than the tail of the respective  Weibull distribution ---a feature  useful for describing the statistics of earthquake
return intervals. The transition point depends on the system size and
the Weibull modulus.

Recent studies have identified
a slope change in logarithmic plots of the ERI pdf, attributed to
spatial (for earthquakes) or temporal (for lab fracturing experiments) non-stationarity of  the background productivity rate~\cite{Baro13}.
We demonstrated that finite-size  effects have a similar impact on the ERI pdf.
Hence, finite size can explain deviations of earthquake return intervals from Weibull scaling without
invoking  non-stationarity (spatial or temporal)
in the background earthquake productivity rate.

In addition, we show that a distinct
feature of the $\kappa$-Weibull distribution  is the dependence of its hazard rate function:
for $m>1$ it increases with increasing time interval up to a certain
threshold,  followed by a  $\propto 1/\tau$ drop.
This is in contrast with the Weibull hazard rate  for $m>1$ which increases indefinitely.
Therefore, the $\kappa$-Weibull distribution allows for temporal clustering of earthquakes independently of the value
of the Weibull modulus.

The application of the
$\kappa$-Weibull distribution to ERI assumes the following:
\begin{itemize}
\item Statistical stationarity, i.e., uniform ERI distribution parameters
over the spatial and temporal  observation window.
\item Renormalizability of the interacting fault system into
an ensemble comprising a finite number of independent effective RVEs with identical interval scale.
\item  Specific but simple functional form for the RVE survival probability given by~\eqref{eq:R1kappa}.
\end{itemize}

We believe that the $\kappa$-Weibull distribution is also potentially useful  for
modeling the fracture strength of heterogeneous quasibrittle structures. The latter involve a finite number of
RVEs and their fracture strength obeys weakest-link scaling~\cite{Bazant06,Bazant09}.
The connection between  ERI  power-law scaling
 and fracture mechanics  pursued herein and in~\cite{dth12} also requires further research.
Finally, we have used statistical methods (Kolmogorov-Smirnov test, Akaike Information Criterion) to
compare different hypotheses for ERI distributions. For example,
 using the Kolmogorov-Smirnov test we showed that for one sequence of earthquakes the $\kappa$-Weibull,
 Weibull, gamma, and generalized gamma probability models  are acceptable at the $5\%$ level.

 \begin{acknowledgments}
M.\ P.\ Petrakis and D. T. Hristopulos acknowledge support by the project \emph{FIBREBREAK},
 Special Research Fund Account, Technical University of Crete.
The Cretan seismic data were graciously provided by D. Becker, Institute of Geophysics, Hamburg University,
Germany. We would also like to thank an anonymous referee for helpful suggestions.
\end{acknowledgments}

\appendix

\section{Inverse Transform Sampling Method}
\label{App:A}
\renewcommand{\theequation}{A-\arabic{equation}}

We generate random numbers from the $\kappa$-Weibull   and the generalized gamma distributions
using the inverse transform sampling method. We first illustrate the algorithm for the
$\kappa$-Weibull random numbers.
\begin{enumerate}
\item  We generate uniform random numbers $u_{\tau} \overset{d}=U(0,1)$.
\item  We employ
the conservation of probability under the variable transformation
$\tau \overset{g(\tau)}\ra u_{\tau}$,  i.e.,
\begin{equation}
\label{eq:rn-trans}
F_{n}(\tau)=F_{U}(u_{\tau})=u_{\tau} \Rightarrow \tau = F^{-1}_{n}(u_{\tau}),
\end{equation}
where
$F_{n}(\tau)= \left( \sqrt{1 + a^{2}\tau^{2m}/n^2} - a\tau^{m}/n \right)^{n} $ and  $a=\tau_{s}^{-m}$.

\item The above in light of~\eqref{eq:rn-trans} leads to
\begin{equation}
\label{eq:rn-kappaW}
\tau = F^{-1}_{n}(u_{\tau})=\left(\frac{n}{2}\right)^{1/m}\, \tau_{s} \,\left( u^{-1/n}_{\tau} - u^{1/n}_{\tau}  \right)^{1/m}
= \tau_{s} \, \left[-\ln_{\kappa}(u_\tau)\right]^{1/m}, \quad \kappa=n^{-1}.
\end{equation}
\end{enumerate}

In the case of the generalized gamma, the cumulative probability distribution is given by
\begin{equation}
\label{eq:Fn-gen-gama}
F_{n}(\tau)= \gamma\left( \frac{k}{m},(\tau/\tau_{s})^m \right),
\end{equation}
where $\gamma(\alpha,x)$ is the \emph{incomplete gamma function} defined by
\[
\gamma(\alpha,x) = \frac{1}{\Gamma(\alpha)}\int_{0}^{x} dy\,\frac{y^{\alpha -1}}{\tau_{s}^{\alpha}}
\, {\rm e}^{-y/\tau_{s}}.
\]
The random numbers $\tau$ are then given by inverting $ \gamma(\alpha, x(\tau)) = u_{\tau}$,
that is by
\[
\tau = \tau_{s} \, \left[\gamma^{-1}_{n}(\alpha; u_{\tau})\right]^{1/m}.
\]

%


\end{document}